\documentclass[pre,twocolumn,superscriptaddress]{revtex4}
\usepackage{graphicx,epsfig}
\usepackage{epstopdf}

\DeclareGraphicsExtensions{. jpg,. pdf, . mps, .png, .eps, . ps, . EPS}

\usepackage{amsmath}
\usepackage{color}
\begin{document}
\newcommand{\beq}{\begin{equation}}
\newcommand{\eeq}{\end{equation}}
\newcommand{\bea}{\begin{eqnarray}}
\newcommand{\eea}{\end{eqnarray}}
\newcommand{\gt}{\tilde{g}}
\newcommand{\mt}{\tilde{\mu}}
\newcommand{\et}{\tilde{\varepsilon}}
\newcommand{\ct}{\tilde{C}}
\newcommand{\bt}{\tilde{\beta}}

\newcommand{\avg}[1]{\langle{#1}\rangle}
\newcommand{\Avg}[1]{\left\langle{#1}\right\rangle}
\newcommand{\cor}[1]{\textcolor{red}{#1}}

\title{Correlations between weights and overlap in ensembles of weighted multiplex networks }
\author{Giulia Menichetti}
\affiliation{Department of Physics and Astronomy and INFN Sez. Bologna, Bologna University, Viale B. Pichat 6/2 40127 Bologna, Italy}
\author{Daniel Remondini}
\affiliation{Department of Physics and Astronomy and INFN Sez. Bologna, Bologna University, Viale B. Pichat 6/2 40127 Bologna, Italy}
\author{Ginestra Bianconi}
\affiliation{School of Mathematical Sciences, Queen Mary University of London, London E1 4NS, United Kingdom}
\begin{abstract}
Multiplex networks describe a large number of systems ranging from social networks to the brain. These multilayer structure encode information in their structure. This information  can be extracted by measuring the  correlations present  in the multiplex networks structure, such as the overlap of the links in different layers. Many multiplex networks are also weighted, and the weights of the links can be strongly correlated with the structural properties of the multiplex network. For example in multiplex network formed by the citation and collaboration networks between  PRE scientists it was found that the statistical properties of citations to co-authors are different from the  one of citations to non-co-authors, i.e. the weights depend on the overlap of the links. Here we present a theoretical framework for modelling multiplex weighted networks with different types of correlations between weights and overlap. To this end, we use the framework of canonical network ensembles, and the recently introduced concept of multilinks,  showing that null models of a large variety of network structures can be constructed in this way.  In order to provide a concrete example of how this framework apply to real data we consider a multiplex constructed from gene expression data of healthy and cancer tissues. 
\end{abstract}
\maketitle

\section{Introduction}
Recently, multilayer networks \cite{PhysReport,Kivela} describing systems as different as   social networks \cite{Thurner}, collaboration networks \cite{Menichetti}, transportation networks \cite{Boccaletti, Vito} climate networks \cite{Kurths} or the brain \cite{Bullmore,GDneuro} are attracting large interest. In fact it has become clear that in order to understand the complexity of a large variety of systems is not enough to consider single networks, but it is necessary to describe  the complex set of interactions between different networks by adopting the framework of multilayer networks. 
For example, the biological functionality of the cells can be described by a multilayer network involving at least metabolic, protein interaction and transcription network layers.
Similarly, social networks cannot be fully understood if the nature of the different ties is not taken into account distinguishing between friendship, collaboration, family ties etc.
Multilayer networks are formed by a set $M$ of layers constituted by single networks, and by interlinks linking the nodes in the different layers. Multilayer networks can be distinguished in multiplex networks \cite{Thurner,Menichetti,Boccaletti,Vito,Kurths} and interacting networks of networks \cite{Havlin3,Netonet}. In interacting networks of networks the nodes in the different layers represent different elements of the system. 
For example, in the cell, metabolites, proteins and transcription factors remain distinct biological entities.
In a multiplex, instead, the same set of nodes forms $M$ networks, one in each layer corresponding to different types of interactions. Examples of multiplex networks are social networks \cite{Thurner} where people can interact in different ways, transportation \cite{Boccaletti,Vito} networks where the same location can be reached by different means of transportation, or collaboration networks \cite{Menichetti,Vito}. 
Here we will provide a multilayer network analysis of a gene network  extracted  using the gene expression of a pool of cancer patients and a pool of healthy subjects respectively for each layer. 

Recently large attention has been given to multiplex network structure \cite{Thurner,Menichetti,Boccaletti,Vito,Kurths,Mucha,PRL,PRE,Goh, Battiston,Spatial,Nonlinear,Kertesz} and dynamics \cite{Havlin1,Diffusion, Radicchi,Perc}.
In particular it has been found that multiplex networks encode in their structure important correlations: we can distinguish for example between degree correlations \cite{Goh,PRL,Nonlinear} determining whether a hub in a network is also an hub in another network, overlap determining to what extent any two nodes of the network are linked in several networks at the same time \cite{Thurner,Boccaletti,PRE,Spatial}, or pairwise activity correlations  measuring if the presence of a node in one network is correlated with the presence of another node in the same network \cite{Vito}. 
Many multiplex networks are also weighted, i.e. the links between the nodes not only are distinguished by the type of interaction linking the nodes, but also by  the intensity of these interactions.
In \cite{Menichetti} different multiplex networks have been extracted from the APS dataset in order to investigate the correlation between the weights of the links and the overlap of the links in different layers. In particular, the multiplex networks formed by the PRE authors in which the scientists are linked if they collaborated with each other and if they cite each other, has been shown to display a  statistical significant difference between the way scientists cite their collaborators and the way scientists cite non-collaborators.
This result shows that in this as in other systems it is possible that the weights of the links are correlated with the pattern of overlap observed between the links of different layers.   
It is therefore very important to propose maximal entropy weighted multiplex networks (based on the theory of  network ensembles \cite{Newman1, Newman2, entropyEPL,BCoolen,AB2009,Coolen,Garlaschelli,uno,due,tre,Multiedge,Multiedge2,delGenio,Weighted09}) models that can be used to generate multiplex networks with different types of correlations. These models, on one side can be used to simulate dynamical processes on different multiplex network topologies, on the other side, similarly to what happens for single networks, their entropy  \cite{entropyEPL,BCoolen}  can be used to evaluate the information content of some of their properties \cite{PNAS,Menichetti}.  
Here we provide the theoretical framework to generate null models for these weighted multiplex networks, using the combined tools of canonical network models (exponential random graphs) and the recently introduced concept \cite{PRE} of multilinks, that is able to   distinguish between different patterns of overlap of the links in the multiplex network. In fact in order to reveal the correlations between the weights distribution and overlap of the links   it is fundamental to consider the weighted properties of the multilinks indicated by the multistrength and inverse multi participation ratio. The multilinks enumerate exhaustively all the types of connections  between two nodes of a multiplex network. Therefore the total number of possible mutlilinks grows exponentially with the number of layers $M$. For this reason the full mutlilink characterization of a multiplex network is numerically feasible only if  the number of layers $M$ is finite.
To overcome this shortcoming here we define the $\nu-$multilinks, that are only characterized by their overlap multiplicity $\nu$, i.e. the $\nu$-multilinks are all the multilinks that connects two nodes of the multiplex with $\nu$ links in $\nu$ different layers. 
By building weighted multiplex ensembles with given properties of the $\nu$-multilinks allows the description and the realization of multiplex networks with large number of layers $M$.    

The paper is structured as follows. In Section II we introduce the weighted multiplex networks and their weighted multilinks properties, in Section III we describe an application to real transcriptomics data, in section IV we introduce weighted multiplex  network ensembles, in section  section V we provide the description of the most relevant weighted multiplex ensembles, considering  the case of uncorrelated and correlated ensembles,  in section VI we show how this framework can be applied to construct null models of the  biological case study, finally in section VII we give the conclusions.

\section{Weighted Multiplex Networks}
\subsection{Definition }
A weighted multiplex is formed by $N$ nodes connected by $M$ weighted networks $G_{\alpha}$, with $\alpha=1,\ldots,M$. A multiplex can be represented as  $\vec{G}=(G_1,G_2,\ldots, G_{\alpha},\ldots G_M)$ where each network $G_{\alpha}$ is fully described by the weighted adjacency matrix of elements $a_{ij}^{\alpha}$, with $a_{ij}^{\alpha}>0$ if there is a link of weight $a_{ij}^{\alpha}$ between node $i$ and node $j$ in layer $\alpha$, otherwise we have $a_{ij}^{\alpha}=0$.\\
In order to simplify the treatment of the weighted multiplex, we suppose that the weight of the link between any pair of nodes $(i,j)$,  $a_{ij}^{\alpha}$ can only assume integer values. This is a legitimate assumption because in a large number of weighted multiplexes  the weights of the links can be considered as multiples of a minimal weight. 
Moreover, for the sake of simplicity we consider only networks without tadpoles and with a symmetric adjacency matrix $\{a_{ij}^{\alpha}\}$, i.e. undirected networks. The generalisation of our approach to directed multiplex networks is straightforward.\\
Since each layer of the multiplex is a weighted network, we can introduce the so-called {\em total strength}, $S_{\alpha}$ that takes into account the total weight of the links in layer $\alpha$. The expression for $S_{\alpha}$ is
\bea
S_{\alpha}=\sum_{i<j} a_{ij}^{\alpha}.
\eea
\subsection{Interaction between the weights and the  topology of single layers}
Each single layer $\alpha$ of the multiplex network is a weighted network \cite{Barrat_PNAS,Aalmas}, namely, a network with heterogeneous interactions between the nodes, that can show interesting weights-topology correlations.  
These correlations can be revealed by measuring the following three quantities:
\begin{itemize}
\item the degree $k_i^{\alpha}$ of a node $i$ in layer $\alpha$,
\item the strength $s_i^{\alpha}$ of node $i$ in layer $\alpha$;
\item the inverse participation ratio $Y_i^{\alpha}$ of node $i$ in layer $\alpha$.
\end{itemize}
These quantities can be expressed in terms of the adjacency matrix elements respectively as
\bea
k_i^{\alpha}=\sum_{j \ne i} \theta(a_{ij}^{\alpha}),
\eea
where the function $\theta(x)=1$ if $x>0$ otherwise $\theta(x)=0$;\\
\bea
s_i^{\alpha}&=&\sum_{j\ne i} a_{ij}^{\alpha}, 
\eea
and 
\bea
Y_i^{\alpha}&=&\sum_{j \ne i} \left(\frac{a_{ij}^{\alpha}}{s_i^{\alpha}}\right)^2.
\eea
Moreover here we introduce for further convenience the quantity $u_i^{\alpha}$ 
\bea
u_i^{\alpha}=Y_i^{\alpha}(s_i^{\alpha})^2=\sum_{j \ne i}\left(a_{ij}^{\alpha}\right)^2,
\eea 
which indicates the sum of the squares of the weights incident to a node.
Similarly to what happens for single networks \cite{Barrat_PNAS,Aalmas}, in  any given layer $\alpha$, the strength $s_i^{\alpha}$ of a node indicates  the sum of the weights of the links of node $i$ in layer $\alpha$, while the inverse participation ratio $Y_i^{\alpha}$ indicates how unevenly the weights of the links of node $i$ in layer $\alpha$ are distributed. The inverse of $Y_i^{\alpha}$ has a range between 1 and $k_i^{\alpha}$. The extremes of the interval correspond respectively to an uniform weight distribution across the links of the node $i$ in the layer $\alpha$, i.e. $a_{ij}^{\alpha}=s_i^{\alpha}/k_i^{\alpha}$, that means $(Y_i^{\alpha})^{-1}=k_i^{\alpha}$, and to the opposite situation, i.e. $(Y_i^{\alpha})^{-1}\approx 1$, when one particular link of the node $i$ has a prevailing weight, i.e. $a_{ir}^{\alpha}\gg a_{ij}^{\alpha}$ for every $j\neq r$. In these terms $Y_i^{\alpha}$ characterises the effective number of links of node $i$ in layer $\alpha$.\\
It is a standard procedure in network theory to evaluate the averages of the strength and the partition ratio of the weights of the links conditioning on the degree of the node. In a multiplex, we will then consider the following quantities 
\bea
s_{\alpha}(k)&=&\Avg{ s_i^{\alpha} \delta(k_i^{\alpha},k)}=\frac{1}{N_k^{\alpha}}\sum_i s_i^{\alpha} \delta(k_i^{\alpha},k)\nonumber \\
Y_{\alpha}(k)&=&\Avg{ Y_i^{\alpha} \delta(k_i^{\alpha},k)}=\frac{1}{N_k^{\alpha}}\sum_i Y_{i,\alpha} \delta(k_i^{\alpha},k)
\eea
where $N_k^{\alpha}$ indicates the number of nodes of degree $k$ in layer $\alpha$.
When considering  $s_k^{\alpha}$, similarly to what happens in general on single networks, we can expect   a scaling of the type
\bea s_{\alpha}(k)\propto k^{\beta_{\alpha}},
\eea 
with $\beta_{\alpha}\ge 1$. We can distinguish \cite{Barrat_PNAS} between two main scenarios depending on the value of the exponent. For $\beta_{\alpha}=1$ the average strength of nodes of degree $k$ increases linearly with $k$. This means that the  average  weight of the links incident to a node does not depend on the degree of the node, at least if we consider only distinguishable links ( for a treatment of the case of   undistinguishable links  see \cite{Multiedge,Multiedge2}).  For $\beta_{\alpha}>1$  hubs tend to have in average links with greater weight than low connectivity nodes.
In a multiplex, we might have that the weights in the different layers are distributed differently.
Therefore we might observe in some layers a superlinear growth of the $s_{\alpha}(k)$ with the degree in that layer, while in other layers we can observe a linear dependence of the strengths on the degree.
When considering single weighted networks it has been observed that in many cases  the inverse participation ratio scales as an inverse power-law of the degree  of the node \cite{Aalmas}.
In the multiplex scenario, this would imply 
\bea
Y_{\alpha}(k)\propto \frac{1}{k^{\xi_{\alpha}}},
\eea
where the exponent $\xi_{\alpha}\leq 1$ might change from one layer to another layer.
The exponent  $\xi_{\alpha}=1$ indicates that all the weights incident to any node are equal, while the exponent $\xi_{\alpha}=0$ would imply the opposite scenario where for every node, one of the weights incident to them is significantly higher than the other weights.

\begin{figure}
\begin{center}
\centerline{\includegraphics[width=3.8 in]{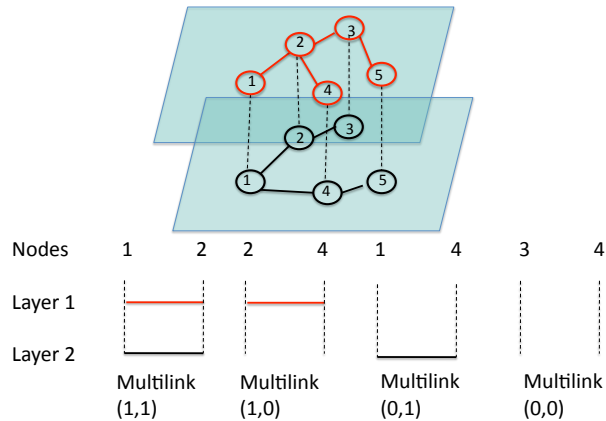}    }
\end{center}
\caption{Schematic view of a duplex (multiplex formed with two networks where any pair of nodes is linked by a different multilink $\vec{m}$.}
\label{multilinkfig}
\end{figure}
\subsection{Weights-topology correlations in multiplex networks with overlap:multilink $\vec{m}$, multistrength $\vec{m}$, and inverse multi partition ratio $\vec{m}$}
It has been recently shown \cite{PRE} that {\em multilinks} are the most natural way to describe and generate    multiplex networks with overlap of the links. 
We say that two nodes are connected by a multilink $\vec{m}=(m_1, m_2,\ldots, m_{\alpha},\ldots, m_M)$ with $m_{\alpha}=0,1$ if they are connected in  every layer $\alpha$   such that $m_{\alpha}=1$ and not connected in every layer $\alpha$ where $m_{\alpha}=0$.  
In figure $\ref{multilinkfig}$ we show an example of a multiplex formed by two layers where each pair of node is linked by a given multilink.
 In order to indicate if a mutlilink $\vec{m}$ is present or not between two given nodes $i$ and $j$ we  can  introduce a  multiadjacency matrix ${\bf A}^{\vec{m}}$ with elements $A^{\vec{m}}_{ij}$ equal to 1 if there is a multilink $\vec{m}$ between node $i$ and node $j$ and zero otherwise.\\
In terms of the weighted adjacency matrices ${\bf a}^{\alpha}$ of the multiplex
 the elements $A^{\vec{m}}_{ij}$ of the multiadjacency matrix ${\bf A}^{\vec{m}}$ are given by 
\begin{equation}
A^{\vec{m}}_{ij}=\prod_{\alpha=1}^{M}[\theta(a_{ij}^{\alpha})m_{\alpha}+ (1-\theta(a_{ij}^{\alpha}))(1-m_{\alpha})]
\label{multilink}
\end{equation}
where $\theta(x)=1$ if $x>0$, otherwise $\theta(x)=0$.
The multilink $\vec{m}=\vec{0}$ between two nodes represents the situation in which in all the layers of the multiplex the two nodes are not directly linked.\\
The multiadjacency matrices are $2^M$ but there are only $2^M-1$ independent multiadjacency matrices because the   normalisation condition
\begin{equation}
\sum_{\vec{m}}A_{ij}^{\vec{m}}=1,
\end{equation}
 is   satisfied for any pair of nodes $(i,j)$. Furthermore, since the multiadjacency matrices have elements $A_{ij}^{\vec{m}}=0,1$, the above condition implies that between any pair of nodes $(i,j)$ there can be only one multilink $\vec{m}$. We indicate the type of this multilink as 
 \bea\vec{m}=\vec{m}^{ij}=(\theta(a_{ij}^1),\theta(a_{ij}^2),\ldots, \theta(a_{ij}^{\alpha}),\ldots,\theta(a_{ij}^M) ),
 \eea where $\theta(x)=1$ if $x>0$ and otherwise $\theta(x)=0$.
 The multilink $\vec{m}$ is characterised by the {\em overlap multiplicity} $\nu(\vec{m})=\sum_{\alpha}m_{\alpha}$ indicating that the  multilink $\vec{m}$ links two pair of nodes by $\nu(\vec{m})$ links.
Using the multiadjacency matrices it is possible to define  the {\em multidegree }$\vec{m}$,  $k_i^{\vec{m}}$ of node $i$, given by 
\bea
k_i^{\vec{m}}=\sum_{j\ne i}A_{ij}^{\vec{m}},
\eea
indicating  how many multilinks $\vec{m}$ are connected to node $i$. Consider for example the social multiplex network where people interact by two means of communication (mobile-phone, email). The multidegree  $k^{(1,1)}_i$ indicates the number of friends of node $i$ that communicate with node $i$ both by email and mobile phone, $k^{(1,0)}_i$ indicates the number of friends of node $i$ that only communicate with node $i$ by mobile-phone and $k^{(0,1)}_i$ indicates the number of friends of node $i$ that only communicate with node $i$ by email.\\
For a given weighted multiplex network we can study the relation between weights and multilinks introducing, at first, the {\em total multistrength} $\vec{m}$, $S^{\vec{m}}_{\alpha}$ in a layer $\alpha$ such that $m_{\alpha}>0$ as 
\bea
S^{\vec{m}}_{\alpha}=\sum_{i<j} a_{ij}^{\alpha}A_{ij}^{\vec{m}}.
\eea
Given a particular multilink $\vec{m}$,  this quantity indicates the total weight in layer $\alpha$ of multilinks $\vec{m}$ and it is properly defined whenever $m_{\alpha}>0$. The  number of total multistrengths $\vec{m}$ that we can define in a multiplex of $M$ layers is given by   $K=M2^{M-1}$. In fact  we have that the total multistrength $S^{\vec{m}}_{\alpha}$ is non-trivial only for multilinks $\vec{m}$ where $m_{\alpha}=1$, while for the remaining layers $\beta$ the value of $m_{\beta}$ can be either zero or one.\\
Moreover we can  define the {\em multistrength} $\vec{m}$, $s^{\vec{m}}_{i,\alpha}$ of node $i$  in layer $\alpha$ such that $m_{\alpha}>0$,   as 
\bea
s^{\vec{m}}_{i,\alpha}=\sum_{j\ne i} a_{ij}^{\alpha}A_{ij}^{\vec{m}}
\eea
and  the {\em inverse  multi participation ratio} $\vec{m}$, $Y^{\vec{m}}_{i,\alpha}$ of node $i$ in layer $\alpha$ such that $m_{\alpha}>0$  as 
\bea
Y^{\vec{m}}_{i,\alpha}=\sum_{j\ne i} \left(\frac{a_{ij}^{\alpha}A_{ij}^{\vec{m}}}{\sum_r a_{ir}^{\alpha} A_{ir}^{\vec{m}}}\right)^2.
\eea
Using the same argument used to evaluate the number of total multistrengths $\vec{m}$, it is easy to prove that the   number of  local multistrength $\vec{m}$ and the number of multi participation ratio $\vec{m}$ are  given by $NM2^{M-1}$.
Moreover here we introduce $u_i^{\alpha,\vec{m}}$, the sum of the squares  of the weights incident  to a node $i$ in layer $\alpha$ and  belonging to a certain type of multilink, as  \bea
u_{i,\alpha}^{ \vec{m}}=Y_{i,\alpha}^{\vec{m}}(s_{i,\alpha}^{\vec{m}})^2=\sum_{j \ne i}\left(a_{ij}^{\alpha}A_{ij}^{\vec{m}}\right)^2.
\eea 

In multiplex weighted networks, it was found that multistrengths and inverse multi partition ratio can have a different scaling behavior depending on the type of multilink. 
In fact  the  average quantities 
\bea
s_{\alpha}^{\vec{m}}(k^{\vec{m}})&=&\Avg{ s_i^{\alpha,\vec{m}} \delta(k_i^{\vec{m}},k^{\vec{m}})}\nonumber \\
Y_{\alpha}^{\vec{m}}(k^{\vec{m}})&=&\Avg{ Y_i^{\alpha,\vec{m}} \delta(k_i^{\alpha,\vec{m}},k^{\vec{m}})}
\eea
are expected to scale like
\bea s_{\alpha}^{\vec{m}}(k^{\vec{m}})&\propto& (k^{\vec{m}})^{\beta_{\alpha,\vec{m}}},\nonumber \\
Y_{\alpha}^{\vec{m}}(k^{\vec{m}})&\propto&(k^{\vec{m}})^{-\xi_{\alpha,\vec{m}}}
\label{scalingm}
\eea 
with  $\beta_{\alpha,\vec{m}}\ge 1$ and positive $\xi_{\alpha,\vec{m}}\leq 1$. The significance dependence of these exponents as a function of the multilink type $\vec{m}$, i.e. on the presence of a certain pattern of overlap or absence of it, indicates the rich interplay between the topology of the weighted networks and their weights.
For example in the CoCi-PRE duplex described in \cite{Menichetti}, formed by authors of PRE that in one layer are connected by collaborations and on the other layer are connected by citations of each other work, the weight-topology correlation is revealed by the different exponent of the multistrength in the citation network calculated either in presence of the overlap of the links in the two layers on in absence of it.This reveals the tendency of   scientific authors of PRE  to cite more the scientists  of high multidegree that are their co-authors than the scientists  with the same multidegree that are not their co-authors. These correlations between weights and overlap patterns are a very general type of correlation likely to exist in large set of multiplex dataset with significant overlap of the links.
It is therefore very important to be able to construct 
null models for multiplex networks with the desired level of correlations between weights and overlap of the links, i.e. with given weighted properties of the multilinks. 
\subsection{Weights-topology correlations in multiplex networks with overlap: $\nu$-total strength,  the $\nu$-multistrength sequence and the $\nu$-inverse multi participation ratio }
Using multilinks $\vec{m}$ can be numerically viable only for
 weighted multiplex networks with a   number $M$ of layers such that $M\ll \log(N)$. As long as this condition is not met,  it is more efficient   to study 
the properties of the $\nu$-multilinks.
The $\nu$-multilinks are any type of mutlilink $\vec{m}$ with multiplicity of overlap $\nu(\vec{m})=\nu$. Therefore in a multiplex social networks, where the layers correspond to the means of communication between two people, node $i$ and node $j$ are linked by a $\nu$-multilink if they can communicate by a maximum of $\nu$ means of communication, independently on the identity of these. For example two people that communicate in Twitter and Facebook are linked by a $\nu$-multilink with $\nu=2$, and the same is true for two people interacting by mobile phone and email.

We can therefore define the $\nu$-multiadjacency matrices ${\bf A}^{\nu}$ with elements $A_{ij}^{\nu}=0,1$ given by 
\bea
A^{\nu}_{ij}&=&\sum_{\vec{m}|\nu(\vec{m})=\nu}A_{ij}^{\vec{m}}
\nonumber \\
&&\hspace*{-15mm}=\sum_{\vec{m}|\nu(\vec{m})=\nu}\prod_{\alpha=1}^{M}[\theta(a_{ij}^{\alpha})m_{\alpha}+ (1-\theta(a_{ij}^{\alpha}))(1-m_{\alpha})],\nonumber
\eea
and $\nu=0,1,2\ldots, M$.
The $\nu$-adjacency matrices are not all independent, since between any two nodes there can be just one type of $\nu$-mutlilink, i.e.
\bea
\sum_{\nu=0}^M A_{ij}^{\nu}=1.
\eea
Therefore we can consider as independent variables only the $\nu$-adjacency matrices corresponding to the non trivial $\nu$-multilinks with $\nu=1,2\ldots, M$.
Moreover we call with $\nu^{ij}$ the type of $\nu$-multilink connecting node $i$ with node $j$, i.e. we have 
\bea
A_{ij}^{\nu^{ij}}=1
\eea
for all pairs of nodes $(i,j)$.
The  number of distinct and non trivial $\nu$-multilinks with $\nu\neq 0$ is given by  $M$, hence the $\nu$-properties of the networks are only polynomial with $M$ while the full mutlilink properties are growing exponentially with $M$. Modelling networks with given $\nu$-mutlilinks properties is therefore convenient when considering multiplex networks with large number of layers $M$.
Given the definition of $\nu$-multiadjacency matrices it is  straightforward to define the  $\nu$-multidegree $k_i^{\nu}$ of node $i$,   given by 
\bea
k_i^{\nu}&=&\sum_{j=1}^N A_{ij}^{\nu}
\eea
indicating the number of neighbors of node $i$ that are connected to node $i$ by a $\nu$-multilink, with $\nu=0,1,2\ldots, M$.
If we consider the weighted properties of the $\nu$-multilink for a given layer $\alpha$, we can define the $\nu$-$total$ $strength$ $S^{\nu}_{\alpha}$,  the $\nu$-$multistrength$ $sequence$ $\{s_{i,\alpha}^{\nu}\}$and the $\nu$-$inverse$ $multi$ $participation$ $ratio$ $\left\{Y_{i,\alpha}^{\nu}\right\}$, as in the following,
\bea
S^{\nu}_{\alpha}&=&\sum_{i<j}a_{ij}^{\alpha}A_{ij}^{\nu}\\
s_{i,\alpha}^{\nu}&=&\sum_{j\neq i} a_{ij}^{\alpha}A_{ij}^{\nu}\nonumber \\
Y_{i,\alpha}^{\nu}&=&\sum_{j\neq i}\left(\frac{a_{ij}^{\alpha}A_{ij}^{\nu}}{\sum_r a_{ir}^{\alpha}A_{ir}^{\nu}}\right)^2.
\eea  

Moreover,  we can introduce the quantities $u_i^{\alpha,\nu}$, indicating  the sum of the squares  of the weights incident to a node $i$ in layer $\alpha$ and  belonging to a certain type of $\nu$-multilink, as  \bea
u_{i,\alpha}^{ \nu}=Y_{i,\alpha}^{\nu}(s_{i,\alpha}^{\nu})^2=\sum_{j \ne i}\left(a_{ij}^{\alpha}A_{ij}^{\nu}\right)^2.
\eea 
Similarly to what described in the previous paragraph, we can evaluate the correlations between the weights and the pattern of overlap between the links by measuring the exponents $\beta_{\alpha,\nu}$ and  $\xi_{\alpha,\nu}$, determining the scaling 
\bea s_{\alpha}^{\nu}(k^{\nu})&\propto& (k^{\nu})^{\beta_{\alpha,\nu}},\nonumber \\
Y_{\alpha}^{\nu}(k^{\nu})&\propto&(k^{\nu})^{-\xi_{\alpha,\nu}}
\label{scalingmnu}
\eea 
of the average quantities $s_{\alpha}^{\nu}(k^{\nu})$ and $Y_{\alpha}^{\nu}(k^{\nu})$ given by  
\bea
s_{\alpha}^{\nu}(k^{\nu})&=&\Avg{ s_i^{\alpha,\nu} \delta(k_i^{\nu},k^{\nu})}\nonumber \\
Y_{\alpha}^{\nu}(k^{\nu})&=&\Avg{ Y_i^{\alpha,\nu} \delta(k_i^{\alpha,\nu},k^{\nu})}.
\eea

\section{A biological case study}
\label{bio}
Here we analyse a  dataset of gene expression profiles from human cancer and healthy subjects, using the framework of mutlilayer networks.
For this analysis, we construct a duplex based on whole-genome gene expression data, as taken from Geo Omnibus Database (\cite{ncbi},  $GSE4183$ dataset). 
From this dataset, a subset of 2835 genes was chosen, known to have a clear biological role (i.e. belonging to known functional pathways as annotated in the KEGG database \cite{kegg}) and with potential interactions between each other (as annotated in PathwayCommons Protein-Protein Interaction network database, \cite{pathway}).
In one layer, the network is reconstructed from gene expression correlation of $N_N = 8$ normal colon samples, while in  the other layer $N_C = 15$ cancer samples are considered.  
We define as $e^{\alpha}_{ij}$ the gene expression value for layer $\alpha$ (normal N or cancer C), in which $i$ is the gene index (ranging from $1$ to $2835$) and $j$ refers to the sample (ranging from $1$ to $8$ for normal samples dataset, and from $1$ to $15$ for cancer samples dataset).\\
Nonparametric Kendall's $\tau$ is used in order to evaluate the correlation between genes, and in each layer a network is obtained by a thresholding on the absolute value of $\tau$ that keeps about $\approx 10\%$ of the possible links ($\tau_1 =0.5$ and $\tau_2 =0.4$ for normal and cancer samples respectively).\\
We also associate a weight $a_{ij}^{\alpha}$ to each duplex link, obtained from gene expression values of the normal and cancer groups. We calculate the average value over all samples for each gene in both layers, namely,
\bea
\avg{e_i^{\alpha}}=\frac{1}{N_{\alpha}}\sum_{k=1}^{N_{\alpha}} e_{ik}^{\alpha}
\eea
with $\alpha=N,C$ and define the weights on each layer as the absolute difference between all gene couples
\bea
a_{ij}^{\alpha}=|\avg{e_i^{\alpha}}-\avg{e_j^{\alpha}}| \quad \forall i,j = 1,\dots,2835.
\eea 
The weights have been discretized as follows: given the minimum and maximum over all values of $a_{ij}^{\alpha}$ (from the union of cancer and normal samples distance matrices), we performed a uniform binning with $100$ bins in this interval, thus obtaining $100$ possible values for the weights $a_{ij}^{\alpha}$. 
This duplex encodes in its topology all the connections among those genes with highly correlated or anticorrelated gene expression profiles. Moreover, the weight distribution describes their distances in terms of mean gene expression values. These kinds of information are essentially different: for example, two genes can be highly correlated in their trends across the samples but one could be much more expressed than the other one.\\
We can integrate different aspects of gene expression data sets thanks to network approaches, and furthermore, we can investigate different experimental setups thanks to multiplex networks tools. The analysis of the multiplex network we have constructed, formed by  one layer for the normal samples  and one layer for those patients with colorectal cancer, can help us understand  if there is a backbone of highly correlated genes that are conserved after the onset of the cancer disease. Moreover, we can characterize all the interactions that are specific for the two conditions.

\begin{figure}
\begin{center}
\centerline{\includegraphics[width=3.8 in]{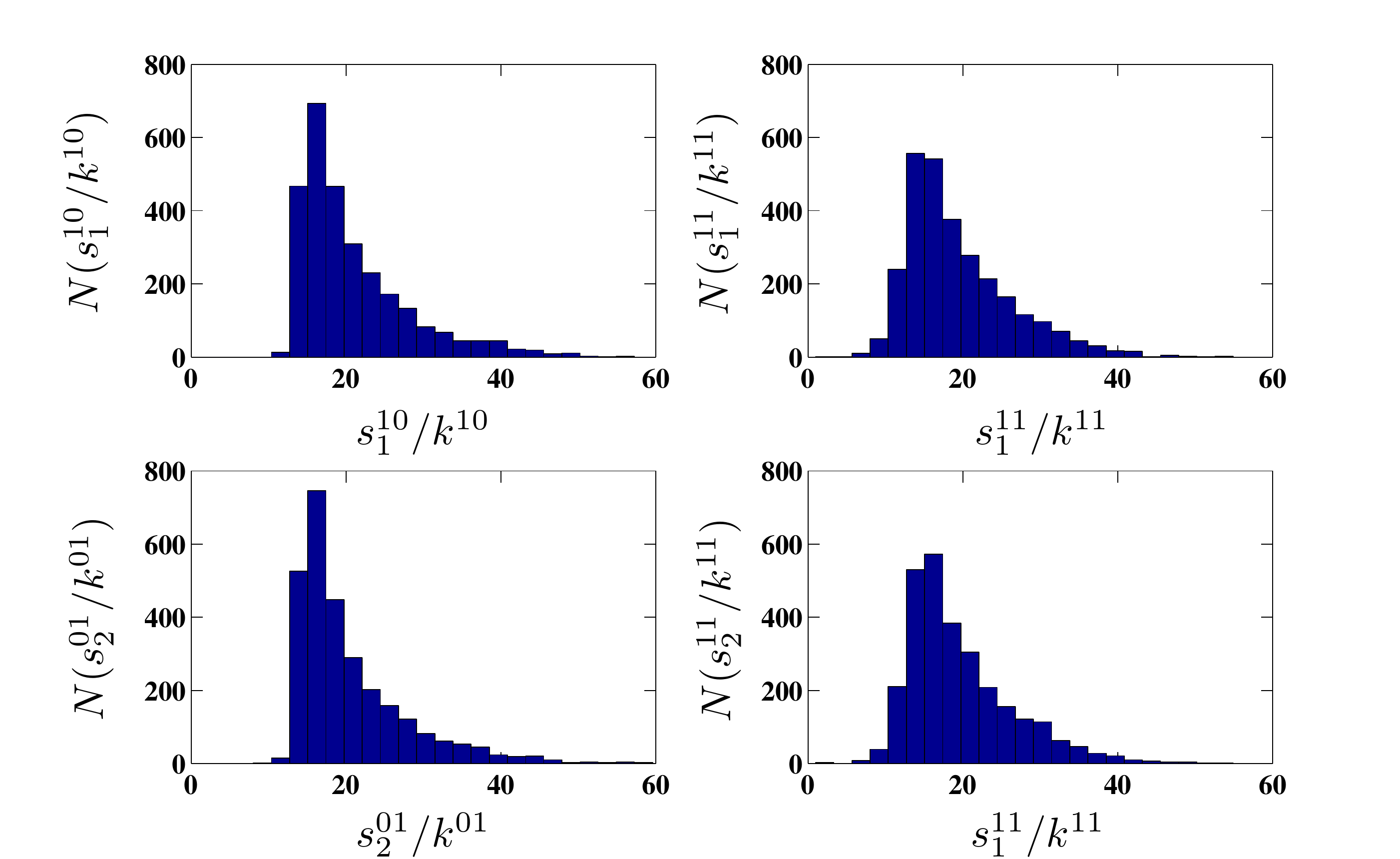}    }
\end{center}
\caption{Biological case study: we display the distributions $\{s_{i, \alpha}^{\vec{m}}/k_{i}^{\vec{m}}\}$, i.e. the average weight of each node's interactions, classified according to the multilinks. }
\label{soverk}
\end{figure}

In order to understand how the weights of the links in a selected layer are related to different multilinks we consider the distributions $\{s_{i, \alpha}^{\vec{m}}/k_{i}^{\vec{m}}\}$, i.e. for each node we calculate the average weight of its interactions, classified according to the multilinks. In Fig. \ref{soverk} we show these distributions, for a given layer $\alpha=1,2$ and a given multilink $\vec{m}$.
In both layers, the distribution of average weights related to multilink $(1,1)$ is significantly different from that one of the specific layer (i.e. multilink $(1,0)$ or $(0,1)$), with a lower mean value and median of the distribution. For layer 1, we compared the distributions $\{s_{i,1}^{(1,1)}/k_{i}^{(1,1)}\}$ and $\{s_{i, 1}^{(1,0)}/k_{i}^{(1,0)}\}$ using a Wilcoxon rank sum test, a nonparametric test for equality of population medians. The p-value is highly significant ($3.88\cdot10^{-22}$) and the two mean values are, respectively,$\Avg{\{s_{i,1}^{(1,1)}/k_{i}^{(1,1)}\}}=19.36$ and $\Avg{\{s_{i,1}^{(1,0)}/k_{i}^{(1,0)}\}}=20.92$. For layer 2, the layer related to cancer samples, the rank sum test is always significant but with a less dramatic p-value ($5.23\cdot10^{-8}$). The mean values for this layer are respectively $\Avg{\{s_{i,2}^{(1,1)}/k_{i}^{(1,1)}\}}=19.54$ and $\Avg{\{s_{i,2}^{(0,1)}/k_{i}^{(0,1)}\}}=20.46$. \\
\newline

We studied the relation between the weights of the set of overlapping links, $\{a_{ij,1}^{(1,1)}\}$ and $\{a_{ij,2}^{(1,1)}\}$. A linear fitting shows that these weights are almost identical, with a relation  $a_{ij,2}^{(1,1)}=0.94\cdot a_{ij,1}^{(1,1)}+3.30$ ($R^2=0.92$).
This result is not trivial, since genes could be correlated (preserving the links) but expressed in a different way (i.e. with different  weights) in healthy and cancer samples, and highlights the existence of a backbone of genes (and related biological processes) that are conserved during the disease progression, possibly due to their fundamental functional role. 


The focus here in this paper is on the possibility to generate a null model for such a multiplex real instance, in order to provide an example of possible application of the theoretical framework here developed to model real datasets. 
In order to generate a null model, we will construct a network ensemble with given multidegree sequence and multistrength sequence and generate multiplex networks out of this ensemble with the desired structural properties. Sampling multiplex networks from their ensembles will offer the opportunity of   comparing our real biological structure with some compatible instances. Moreover, the entropy measure gives us the logarithm of the number of ``typical'' duplex networks in the ensemble, a value that can be used to compare different experimental setups and clinical conditions, evaluating  what is the level of information encoded in the selected structural properties of biological networks.\\

\section{Canonical weighted  multiplexes ensembles or exponential weighted multiplexes}

Null models for weighted multiplex networks can be constructed using the formalism of canonical network ensembles also known as exponential random graphs \cite{Newman1,Newman2,AB2009,Garlaschelli}.
These ensembles of networks  generate the least biased set of networks satisfying a set of constraint on average.
In fact,  these ensembles  are derived by a maximal entropy approach conditioned to a series of structural constraints.
The entropy of these ensembles and of the correspondent microcanonical ensembles enforcing the corresponding hard constraints \cite{entropyEPL,PNAS}, can be used to quantify the level of information encoded in the structural constraints that are imposed to the networks.   
In \cite{PRE, Spatial} this approach was taken to model simple multiplex networks. Here we show how this framework can be applied to model weighted multiplex networks.

A weighted multiplex ensemble is defined once the probability $P(\vec{G})$ of any possible weighted multiplex is given.
We can build a canonical multiplex ensemble by maximizing the entropy ${\cal S}$ of the ensemble given by 
\bea
{\cal S}=-\sum_{\vec{G}}P(\vec{G})\log P(\vec{G})
\label{entropy}
\eea under the condition that the soft constraints we want to impose are satisfied.
We assume to have $K$ of such constraints determined by the conditions
\begin{equation}
\sum_{\vec{G}}P(\vec{G})F_{\mu}(\vec{G})=C_{\mu}
\label{constraints}
\end{equation}
for $\mu=1,2\ldots, K$, where $F_{\mu}(\vec{G})$ determines one of the  structural constraints that we want to impose to the multiplex. Therefore, the maximal-entropy multiplex ensemble satisfying the constraints given by Eqs.~(\ref{constraints}) is the solution of the following system of equations 
\begin{equation}
\frac{\partial }{\partial P(\vec{G})}\left[{\cal S}-\sum_{\mu=1}^K \lambda_{\mu} \sum_{\vec{G}}F_{\mu}(\vec{G})P(\vec{G})-\Lambda\sum_{\vec{G}}P(\vec{G})\right]=0,
\end{equation}
where the Lagrangian multiplier $\Lambda$ enforces the normalisation of the $P(\vec{G})$ probability distribution, and the Lagrangian multiplier $\lambda_{\mu}$ enforces the constraint $\mu$.\\
Therefore we get  that the probability of a multiplex $P(\vec{G})$ in a canonical multiplex ensemble is given by 
\begin{equation}
P(\vec{G})=\frac{1}{Z}\exp\left[-\sum_{\mu}\lambda_{\mu}F_{\mu}(\vec{G})\right]
\label{PC}
\end{equation}
where the normalisation constant $Z=\exp(1+\Lambda)$ is called the  ``partition function " of the canonical multiplex ensemble and is fixed by the normalisation condition on $P(\vec{G})$.
The values of the Lagrangian multipliers $\lambda_{\mu}$ are  determined  by imposing the constraints given by Eq. $(\ref{constraints})$, assuming for the probability $P(\vec{G})$ the structural form given by Eq. $(\ref{PC})$.
From the definition of the partition function $Z$ and Eq. $(\ref{PC})$, it can be easily shown that the Lagrangian multipliers $\lambda_{\mu}$ can be expressed as the solutions of the following set of equations, 
\bea
C_{\mu}=-\frac{\partial \log Z}{\partial \lambda_{\mu}}.
\eea
We call the entropy ${\cal S}$ of the canonical multiplex ensemble the {\it Shannon entropy} of the ensemble.

Further on, we can define the marginal probability for a specific value of the element $a_{ij}^{\alpha}$ as
\begin{equation}
\label{marginals}
\pi_{ij}^{\alpha}(a_{ij}^{\alpha}=w)=\sum_{\vec{G}}P(\vec{G})\delta(a_{ij}^{\alpha}, w)
\end{equation}
where $\delta(x,y)$ stands for the Kronecker delta. The marginal probabilities $\pi_{ij}^{\alpha}(a_{ij}^{\alpha})$ sum up to one
\begin{equation}
\sum_{a_{ij}^{\alpha}=0}^{\infty}\pi_{ij}(a_{ij}^{\alpha})=1
\end{equation}
We can compute also the average weight $\avg{a_{ij}^{\alpha}}$ between node $i$ and node $j$  that is
\bea
\Avg{a_{ij}^{\alpha}}&=&\sum_{\vec{G}}P(\vec{G})a_{ij}^{\alpha}=\sum_{a_{ij}^{\alpha}=0}^{\infty}a_{ij}^{\alpha}\pi_{ij}(a_{ij}^{\alpha})
\label{av_weight}
\eea

In the layer $\alpha$ a link between two nodes $i$ and $j$ exists with probability $p_{ij}^{\alpha}$, that is related with all the possible weights different from zero
\begin{equation}
\label{probability}
p_{ij}^{\alpha}=\sum_{\vec{G}}P(\vec{G}) \theta(a_{ij}^{\alpha})=\sum_{a_{ij}^{\alpha}\ne 0}^{\infty}\pi_{ij}^{\alpha}(a_{ij}^{\alpha}).
\end{equation}

\subsection{Uncorrelated and correlated canonical multiplex ensembles}
The multiplex ensembles can be distinguished between uncorrelated and correlated multiplex ensembles.
For uncorrelated multiplex ensembles, the probability of a multiplex $P(\vec{G})$ is factorizable into the probability $P_{\alpha}(G_{\alpha})$ of each single network  $G_{\alpha}$ at layer $\alpha$, i.e.
\begin{equation}
\label{Puncorr}
P(\vec{G}) =\prod_{\alpha=1}^{M}P_{\alpha}(G_{\alpha}).
\end{equation}
Therefore, the entropy ${\cal S}$ of any uncorrelated multiplex ensemble given by Eq. $(\ref{entropy})$ with $P(\vec{G})$ given by Eq. $(\ref{Puncorr})$ is additive in the number of layers, i.e.
\begin{equation}
\label{entropy_noncorr}
{\cal S}=\sum_{\alpha=1}^M {\cal S}_{\alpha}=-\sum_{\alpha=1}^M \sum_{G^{\alpha}}P_{\alpha}(G_{\alpha})\log P_{\alpha}(G_{\alpha})
\end{equation}
 As a consequence of these relations, when  each constraint depends on a single network $G_{\alpha}$ in a layer $\alpha$ the resulting multiplex ensemble is uncorrelated.\\
Example of these types of constraints are the total strengths $S^{\alpha}$ in each layer $\alpha$, the strength $s_i^{\alpha}$ of the generic node $i$ in layer $\alpha$, or the degree $k_i^{\alpha}$ of the node $i$ in layer $\alpha$.\\
In these ensembles of multiplex networks we have that the presence of a link in a layer $\alpha$ is uncorrelated with the presence of a link between the same two nodes in a layer $\beta\neq \alpha$. Therefore we have 
\bea
\Avg{a_{ij}^{\alpha}a_{ij}^{\beta}}=\Avg{a_{ij}^{\alpha}}\Avg{a_{ij}^{\beta}}.
\eea
In correlated multiplex networks, instead the probability of a multiplex does not factorize into the probabilities of the single networks that constitute the multiplex network.
We have in this case 
\begin{equation}
\label{Pcorr}
P(\vec{G}) \neq \prod_{\alpha=1}^{M}P_{\alpha}(G_{\alpha}).
\end{equation}
and as a consequence of this there is  at least a pair of nodes $(i,j)$ and layers $\alpha,\beta$ such that the weights  of the links connecting node $i$ and node $j$ is layer $\alpha $ and layer  $\beta$ are correlated, i.e.
\bea
\Avg{a_{ij}^{\alpha}a_{ij}^{\beta}}\neq\Avg{a_{ij}^{\alpha}}\Avg{a_{ij}^{\beta}}.
\eea
Example of  constraints that generate correlated multiplex ensembles are constraints on the multidegree sequence or the multistrength sequence.\\

 \section{Examples of correlated and uncorrelated multiplex network ensembles}

 Here we provide three example of uncorrelated and correlated multiplex network ensemble.
 The case of uncorrelated multiplex networks (Case treated in Sec. \ref{1}) is very closely related to the treatment of weighted ensembles of single networks \cite{uno,due,tre}, nevertheless the case of uncorrelated multiplex networks  (Cases treated in Sec. \ref{2}-\ref{3})provides a novel framework to understand  correlations between weights and multidegrees in a model. 
 In the main text of the article we will present only few examples of multiplex network ensembles, while a longer set of ensembles is discussed in the appendix.

\subsubsection{Multiplex ensembles with given expected strength sequence and degree sequence in each layer}
\label{1}
This is an example of uncorrelated network ensemble.
We fix the expected strength $s_i^{\alpha}$ and the expected degree $k_i^{\alpha}$ of every node $i$, in each layer $\alpha$. We have $K=M \cdot 2N$ constraints in the system. These constraints are given by
\begin{align}
\sum_{\vec{G}}F_{i,\alpha}(\vec{G}) P(\vec{G})&=\sum_{\vec{G}}\left ( \sum_{j \ne i} a_{ij}^{\alpha}\right)P(\vec{G})=s_i^{\alpha}\nonumber \nonumber \\
\sum_{\vec{G}}F_{i,\alpha}(\vec{G}) P(\vec{G})&=\sum_{\vec{G}}\left ( \sum_{j \ne i} \theta(a_{ij}^{\alpha})\right)P(\vec{G})=k_i^{\alpha},
\end{align}
with $\alpha=1,2,\ldots, M$.
We introduce the Lagrangian multipliers $\lambda_{i,\alpha}$ for the first set of $N\cdot M$ constraints and the Lagrangian multipliers $\omega_{i,\alpha}$ for the second set of $N\cdot M$ constraints. Therefore, the probability $P(\vec{G})$ of a multiplex in this ensemble, of general expression given by Eq. (\ref{PC}), in this specific example is given by  
\begin{equation}\hspace*{-3mm}
P(\vec{G})=\frac{1}{Z}\exp \left [ -\sum_{\alpha=1}^M \sum_i\lambda_{i,\alpha} \sum_{j \ne i} a_{ij}^{\alpha}-\sum_{\alpha=1}^M \sum_i \omega_{i,\alpha} \sum_{j \ne i} \theta(a_{ij}^{\alpha})\right]\nonumber
\end{equation}
where the partition function $Z$ can be expressed explicitly as
\bea
Z&=&\sum_{\vec{G}}\exp \left [ -\sum_{\alpha=1}^M \sum_i \sum_{j \ne i}\left(\lambda_{i,\alpha}  a_{ij}^{\alpha}+\omega_{i,\alpha} \theta(a_{ij}^{\alpha})\right)\right]\nonumber\\
&=&\prod_{\alpha=1}^{M}\prod_{i<j}\left ( 1+ \frac{e^{-(\omega_{i,\alpha}+\omega_{j,\alpha})-(\lambda_{i,\alpha}+\lambda_{j,\alpha})}}{1-e^{-(\lambda_{i,\alpha}+\lambda_{j,\alpha})}}\right),
\eea
and the Lagrangian multipliers are fixed by the conditions
\begin{align}
s_i^{\alpha}&=-\frac{\partial{\log Z}}{\partial{\lambda_{i,\alpha}}}\nonumber\\
k_i^{\alpha}&=-\frac{\partial{\log Z}}{\partial{\omega_{i,\alpha}}}
\end{align}
The average weight of the link $(i,j)$ in layer $\alpha$, i.e. $\Avg{a_{ij}^{\alpha}}$, is given by   Eq. (\ref{av_weight}) that in this case reads 
\begin{align}
\Avg{a_{ij}^{\alpha}}&=
&=\frac{e^{-(\omega_{i,\alpha}+\omega_{j,\alpha})+(\lambda_{i,\alpha}+\lambda_{j,\alpha})}}{(e^{\lambda_{i,\alpha}+\lambda_{j,\alpha}}-1)(e^{-(\omega_{i,\alpha}+\omega_{j,\alpha})}+e^{\lambda_{i,\alpha}+\lambda_{j,\alpha}}-1)}
\end{align}
From Eq. (\ref{marginals}) we write the marginal probabilities $\pi_{ij}^{\alpha}(a_{ij}^{\alpha})$ for this specific ensemble that is given by 
\begin{align}
\pi_{ij}^{\alpha}(a_{ij}^{\alpha})
&=\frac{e^{-(\lambda_{i,\alpha}+\lambda_{j,\alpha})a_{ij}^{\alpha}-(\omega_{i,\alpha}+\omega_{j,\alpha})\theta(a_{ij}^{\alpha})}(1-e^{-(\lambda_{i,\alpha}+\lambda_{j,\alpha})})}{ 1+ e^{-(\lambda_{i,\alpha}+\lambda_{j,\alpha})}(e^{-(\omega_{i,\alpha}+\omega_{j,\alpha})}-1)}.
\label{pib}
\end{align}

Moreover, from Eq. (\ref{probability}) the probability $p_{ij}^{\alpha}$ that the link $(i,j)$ in layer $\alpha$ has weight different from zero is given by 
\begin{align}
p_{ij}^{\alpha}&=&\frac{e^{-(\omega_{i,\alpha}+\omega_{i,\alpha})}}{e^{-(\omega_{i,\alpha}+\omega_{j,\alpha})}+e^{\lambda_{i,\alpha}+\lambda_{j,\alpha}}-1}
\end{align}
We observe that we can write the Eq. (\ref{PC}) in terms of marginal probabilities $\pi_{ij}^{\alpha}(a_{ij}^{\alpha})$, namely
\begin{equation}
P(\vec{G}) =\prod_{\alpha=1}^{M}\prod_{i<j}\pi_{ij}^{\alpha}(a_{ij}^{\alpha}).
\label{PCl}
\end{equation}
Therefore the  entropy ${\cal S}$ of this canonical multiplex ensemble is given by Eq.~(\ref{entropy}) and in this special case can be written as
\begin{equation}
\label{entropyS}
\mathcal{S}=-\sum_{\alpha=1}^M\sum_{i<j}\sum_{a_{ij}^{\alpha}=0}^{\infty}\pi_{ij}^{\alpha}(a_{ij}^{\alpha})\log (\pi^{\alpha}_{ij}(a_{ij}^{\alpha})).
\end{equation}

\subsubsection{Multiplex ensembles with given expected multidegree sequence $\{k_{i}^{\vec{m}}\}$ and given expected multistrength sequence $\{s_{i, \alpha}^{\vec{m}}\}$}
\label{2}
In many applications it is important to   consider the weighted multiplex networks in which we fix at the same time the average multidegree sequence $k_{i}^{\vec{m}}$ and  the average multistrength sequence $s_{i, \alpha}^{\vec{m}}$. The number of independent constraints is therefore $K=(2^M-1) \cdot N + (2^{M-1})\cdot M \cdot N$.\\
In particular, the constraints we are imposing are the following, 
\begin{align}
\sum_{\vec{G}}F_{i, \alpha}^{\vec{m}}(\vec{G}) P(\vec{G})&=\sum_{\vec{G}}\left ( \sum_{j\ne i}A_{ij}^{\vec{m}}a_{ij}^{\alpha}\right)P(\vec{G})=s_{i, \alpha}^{\vec{m}}\nonumber\\
\sum_{\vec{G}}F_{i}^{\vec{m}}(\vec{G}) P(\vec{G})&=\sum_{\vec{G}}\left ( \sum_{j\ne i}A_{ij}^{\vec{m}}\right)P(\vec{G})=k_{i}^{\vec{m}}.
\end{align}
The canonical probability $P(\vec{G})$ of the multiplex in the ensemble becomes
{
\bea
P(\vec{G})&=&\frac{1}{Z}\exp \left [ -\sum_{\vec{m}\neq \vec{0}}\sum_i \sum_{j \ne i} \left ( \omega_i^{\vec{m}}A_{ij}^{\vec{m}}+
\sum_{\alpha=1}^M \lambda_{i,\alpha}^{\vec{m}} A_{ij}^{\vec{m}} a_{ij}^{\alpha}\right )\right]\nonumber \\
&=&\frac{1}{Z}\exp \left [ -\sum_{i<j}\sum_{\vec{m}\neq \vec{0}}(\omega_i^{\vec{m}}+\omega_j^{\vec{m}})A_{ij}^{\vec{m}}\right]\times\nonumber \\
&\times & \exp\left[-\sum_{i<j} \sum_{\vec{m}\neq\vec{0}}\sum_{\alpha=1}^M (\lambda_{i,\alpha}^{\vec{m}}+\lambda_{j,\alpha}^{\vec{m}})A_{ij}^{\vec{m}} a_{ij}^{\alpha}\right]
\eea
}
The partition function $Z$ can be expressed explicitly as
\begin{align}
Z&=\prod_{i<j}\mathcal{Z}_{ij}
\end{align}
where $\mathcal{Z}_{ij}$ is given by  
\begin{equation}
\mathcal{Z}_{ij}=1+\sum_{\vec{m}\ne \vec{0}}e^{-(\omega_i^{\vec{m}}+\omega_j^{\vec{m}})}\prod_{\alpha=1}^{M}\left ( \frac{e^{-(\lambda_{i,\alpha}^{\vec{m}} +\lambda_{j,\alpha}^{\vec{m}}  )}}{1-e^{-(\lambda_{i,\alpha}^{\vec{m}} +\lambda_{j,\alpha}^{\vec{m}}  )}}\right )^{m_{\alpha}}
\end{equation}

The Lagrangian multipliers are fixed by the conditions
\begin{align}
-\frac{\partial{\log Z}}{\partial{\lambda_{i,\alpha}^{\vec{m}}}}&=s_{i,\alpha}^{\vec{m}}=\sum_{j \ne i}\Avg{a_{ij}^{\alpha}A_{ij}^{\vec{m}}},\nonumber\\
-\frac{\partial{\log Z}}{\partial{\omega_{i}^{\vec{m}}}}&=k_{i}^{\vec{m}}=\sum_{j \ne i}\Avg{A_{ij}^{\vec{m}}}.
\label{Lagrangians}
\end{align}

We now indicate with   $\vec{a}_{ij}$ the vector $(a_{ij}^1,a_{ij}^2,\ldots, a_{ij}^{\alpha},\ldots, a_{ij}^M)$.
The probability of a multiplex $P(\vec{G})$ can be rewritten as
\begin{equation}
P(\vec{G})=\prod_{i<j}\pi_{ij}(\vec{a}_{ij}),
\label{probability_wmultiplex}
\end{equation}
with
\begin{equation}
\pi_{ij}(\vec{a}_{ij})=\frac{e^{-(\omega_i^{\vec{m}^{ij}}+\omega_j^{\vec{m}^{ij}})}}{\mathcal{Z}_{ij}} { e^{-\sum_{\alpha=1,M}(\lambda_{i,\alpha}^{\vec{m}^{ij}}+\lambda_{j,\alpha}^{\vec{m}^{ij}})a^{\alpha}_{ij}}}
\label{pistrengthdegreemulti}
\end{equation}
where $\vec{m}^{ij}=(m^{ij}_1,\ldots, m^{ij}_{\alpha},\ldots, m^{ij}_m)$ with $m^{ij}_{\alpha}=\theta(a_{ij}^{\alpha})$. With $\pi_{ij}(\vec{a}_{ij})$ we define, for a position $ij$,  the probability of a particular sequence of weights on the layers. The normalization condition is fulfilled
\begin{equation}
\sum_{\vec{a}_{ij}}\pi_{ij}(\vec{a}_{ij})=1.
\label{piijnorm}
\end{equation}

Further on we can compute the average weight of the link $ij$ on the multilink $\vec{m}$, in the layer $\alpha$ and the probability of a multilink $\vec{m}$ between node $i$ and node $j$, $p_{ij}^{\vec{m}}=\Avg{A_{ij}^{\vec{m}}}$, respectively,
\bea
\Avg{a_{ij}^{\alpha}A_{ij}^{\vec{m}}}&=&\frac{e^{-(\omega_{i}^{\vec{m}}+\omega_{j}^{\vec{m}})}}{\mathcal{Z}_{ij}} \left ( \frac{1}{1-e^{-(\lambda_{i,\alpha}^{\vec{m}}+\lambda_{j,\alpha}^{\vec{m}})}}\right )\times \nonumber \nonumber \\
&&\times\prod_{\beta=1}^{M}\left ( \frac{e^{-(\lambda_{i,\beta}^{\vec{m}}+\lambda_{j,\beta}^{\vec{m}})}}{1-e^{-(\lambda_{i,\beta}^{\vec{m}}+\lambda_{j,\beta}^{\vec{m}})}}\right )^{m_{\beta}}\\
\hspace*{-3mm}p_{ij}^{\vec{m}}&\hspace*{-3mm}=&\hspace*{-3mm}\frac{e^{-(\omega_i^{\vec{m}}+\omega_j^{\vec{m}})}}{\mathcal{Z}_{ij}}\prod_{\alpha=1}^{M}\left ( \frac{e^{-(\lambda_{i,\alpha}^{\vec{m}} +\lambda_{j,\alpha}^{\vec{m}}  )}}{1-e^{-(\lambda_{i,\alpha}^{\vec{m}} +\lambda_{j,\alpha}^{\vec{m}}  )}}\right )^{m_{\alpha}}
\label{pstrengthdegreemulti}
\eea
where the  normalization condition is fulfilled, namely,
\begin{equation}
\sum_{\vec{m}}p_{ij}^{\vec{m}}=1.
\label{pijnorm}
\end{equation}
Moreover, the relationship between $p_{ij}^{\vec{m}}$ and the probabilities $\pi_{ij}(\vec{a}_{ij})$ is
\begin{equation}
\sum_{\vec{a}_{ij}}A_{ij}^{\vec{m}}\pi_{ij}(\vec{a}_{ij})=p_{ij}^{\vec{m}}.
\label{pijpiij1}
\end{equation}

Finally,  the probability of a multiplex $P(\vec{G})$ is given by Eq.~(\ref{probability_wmultiplex}) and the entropy ${\cal S}$ of this ensemble can be calculated starting from its definition Eq. $(\ref{entropy})$, giving 
\bea
{\cal S}&=-\sum_{i<j}\sum_{\vec{a}_{ij}} \pi_{ij}(\vec{a}_{ij})\log  \pi_{ij}(\vec{a}_{ij}).
\label{entropy2}
\eea

\subsubsection{Multiplex ensembles with given expected $\nu$-multidegree sequence $\{k_{i}^{\nu} \}$ and expected $\nu$-multistrength sequence $\{s_{i,\alpha}^{\nu} \}$ }
\label{3}
In a multiplex networks formed by many layers, an efficient way to consider both topological and weighted properties of the multilayer structure is to construct multiplex networks  with given expected $\nu$-multidegree sequence $\{k_i^{\nu}\}$ and expected $\nu$-multistrength sequence $\{s_{i,\alpha}^{\nu} \}$. The $N\cdot M \cdot (M+1)$ constraints are given by 
\begin{align}
\sum_{\vec{G}}F_{i, \alpha}^{\nu}(\vec{G}) P(\vec{G})&=\sum_{\vec{G}}\left ( \sum_{j\ne i}a_{ij}^{\alpha}A_{ij}^{\nu}\right)P(\vec{G})=s_{i, \alpha}^{\nu}\nonumber\\
\sum_{\vec{G}}F_{i}^{\nu}(\vec{G}) P(\vec{G})&=\sum_{\vec{G}}\left ( \sum_{j\ne i}A_{ij}^{\nu}\right)P(\vec{G})=k_{i}^{\nu},
\label{conditionsnu3}
\end{align}
with $i=1,2,\ldots, N$, $\alpha=1,2,\ldots M$ and $\nu=1,2,\ldots, M$.
The canonical probability $P(\vec{G})$ of the multiplex in this ensemble can be expressed in terms of the Lagrangian multipliers $\lambda_{j,\alpha}^{\nu}$ and $ \omega_i^{\nu}$, i.e. 
\bea
P(\vec{G})&=&\frac{1}{Z}\exp \left [ -\sum_{i<j}\sum_{\nu=1}^{M}(\omega_i^{\nu}+\omega_j^{\nu})A_{ij}^{\nu}\right]\times \\
&&\times \exp\left[-\sum_{i<j} \sum_{\nu=1}^{M}\sum_{\alpha=1}^M (\lambda_{i,\alpha}^{\nu}+\lambda_{j,\alpha}^{\nu})A_{ij}^{\nu} a_{ij}^{\alpha}\right],\nonumber
\eea
where the  partition function $Z$ is given by 
\begin{align}
Z&=\prod_{i<j}\mathcal{Z}_{ij},
\end{align}
with   
\begin{equation}
\mathcal{Z}_{ij}=1+\sum_{\nu=1}^{M}e^{-(\omega_i^{\nu}+\omega_j^{\nu})}\sum_{\vec{m}|\nu(\vec{m})=\nu}\prod_{\alpha=1}^{M}\left ( \frac{e^{-(\lambda_{i,\alpha}^{\nu} +\lambda_{j,\alpha}^{\nu}  )}}{1-e^{-(\lambda_{i,\alpha}^{\nu} +\lambda_{j,\alpha}^{\nu}  )}}\right )^{m_{\alpha}}
\end{equation}

The Lagrangian multipliers are fixed by the conditions Eq.~$(\ref{conditionsnu3})$ that can be also written in terms of the partial derivatives of the partition function as 
\bea
-\frac{\partial{\log Z}}{\partial{\lambda_{i,\alpha}^{\nu}}}&=&s_{i,\alpha}^{\nu}=\sum_{j \ne i}\Avg{a_{ij}^{\alpha}A_{ij}^{\nu}},\nonumber\\
-\frac{\partial{\log Z}}{\partial{\omega_{i}^{\nu}}}&=&k_{i}^{\nu}=\sum_{j \ne i}\Avg{A_{ij}^{\nu}}.
\eea
As in the previous cases, the probability $P(\vec{G})$ of a multiplex network $\vec{G}$ is given by Eq.~$(\ref{probability_wmultiplex})$. The entropy of this ensemble takes the same expression given by Eq.~ $(\ref{entropy2})$ with $\pi_{ij}(\vec{a}_{ij})$ given by 
\bea
\pi_{ij}(\vec{a}_{ij})=\frac{e^{-(\omega_i^{\nu^{ij}}+\omega_j^{\nu^{ij}})}}{\mathcal{Z}_{ij}} { e^{-\sum_{\alpha=1,M}(\lambda_{i,\alpha}^{\nu^{ij}}+\lambda_{j,\alpha}^{\nu^{ij}})a^{\alpha}_{ij}}}
\eea

The probability $p_{ij}^{\nu}$ that the node $i$ and the node $j$ are linked by a $\nu$-multilink is given by 
\bea
p_{ij}^{\nu}&=&\frac{e^{-(\omega_i^{\nu}+\omega_j^{\nu})}}{\mathcal{Z}_{ij}}\sum_{\vec{m}|\nu(\vec{m})=\nu}\prod_{\alpha=1}^{M}\left ( \frac{e^{-(\lambda_{i,\alpha}^{\nu} +\lambda_{j,\alpha}^{\nu}  )}}{1-e^{-(\lambda_{i,\alpha}^{\nu} +\lambda_{j,\alpha}^{\nu}  )}}\right )^{m_{\alpha}}
\eea
Finally, the average weight of the link $a_{ij}^{\alpha}$ belonging to a $\nu$-multilink is given by 
\bea
\Avg{a_{ij}^{\alpha}A_{ij}^{\nu}}&=&\frac{e^{-(\omega_{i}^{\nu}+\omega_{j}^{\nu})}}{\mathcal{Z}_{ij}} \left ( \frac{1}{1-e^{-(\lambda_{i,\alpha}^{\nu}+\lambda_{j,\alpha}^{\nu})}}\right )\times\\
&\times &\sum_{\vec{m}|\nu(\vec{m})=\nu}m_{\alpha}\prod_{\beta=1}^{M}\left ( \frac{e^{-(\lambda_{i,\beta}^{\nu}+\lambda_{j,\beta}^{\nu})}}{1-e^{-(\lambda_{i,\beta}^{\nu}+\lambda_{j,\beta}^{\nu})}}\right )^{m_{\beta}}\nonumber
\eea

\begin{figure*}
\center
\includegraphics[width=\textwidth]{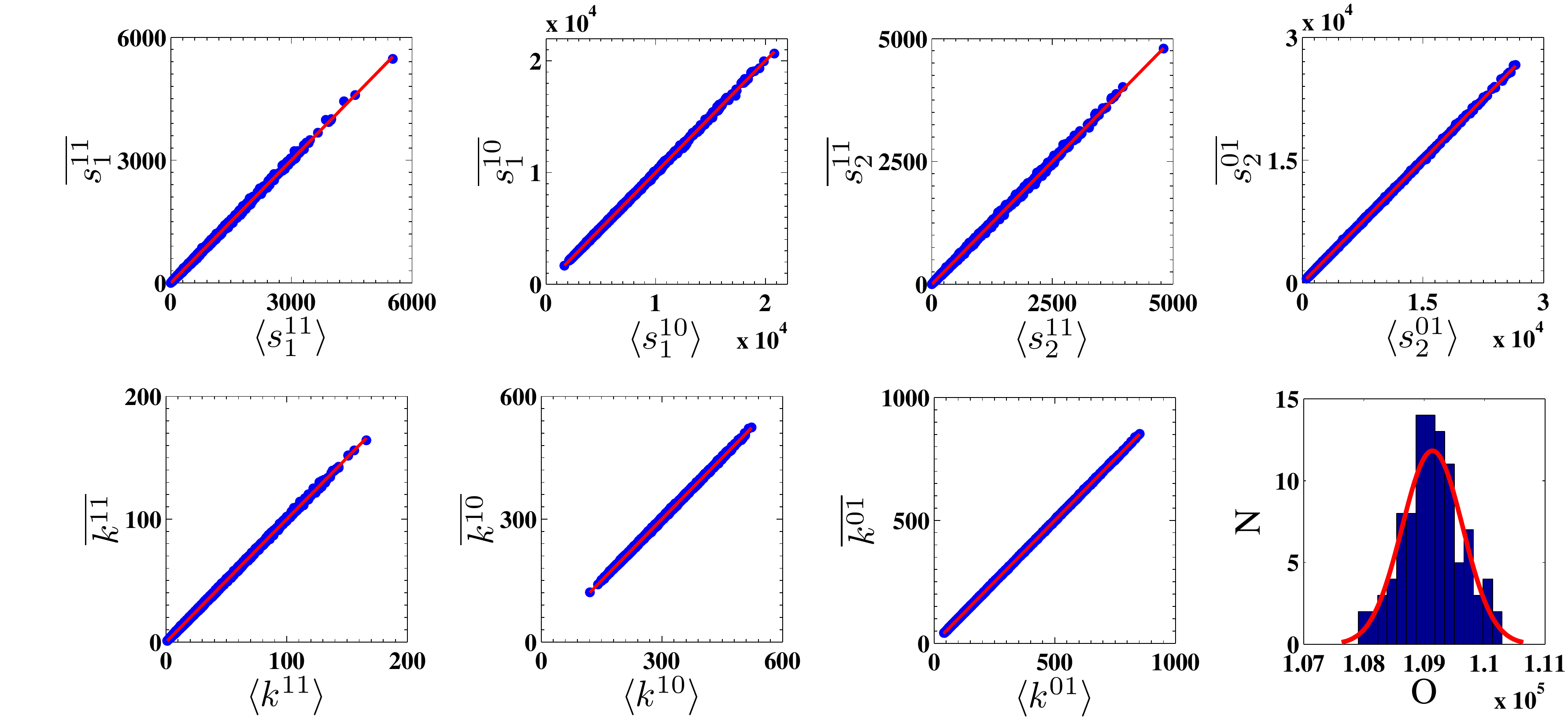}
\caption{Comparison of the real values of multistrength and multidegree sequence with their related average values calculated over  100  instances. Angular brackets $(\Avg{\dots})$ indicate the real values (the fixed average values of the canonical ensemble), while overbar (${\bar{\dots}}$)  defines the average measure over the 100 duplexes. 
Considering the relative error between the real values and the average values for each node, $\Delta E_i = (\bar{x_i} - \langle x_i \rangle)/\langle x_i \rangle$ $i=1,\dots, 2,835$,  the average absolute relative error $ \langle |\Delta E| \rangle$, over all nodes for each measure, ranges from a minimum of $0.5\%$ to a maximum of $2.4\%$.
In the last panel we display the distribution of the $100$ measures of the overlap between the two layers (in the real duplex this value was $109,056$ links). The red line is a Gaussian distribution with the same mean and variance as the empirical distribution.
}
\label{average_ensemble_total}
\end{figure*}


\begin{figure}
\begin{center}
\centerline{\includegraphics[width=3.8 in]{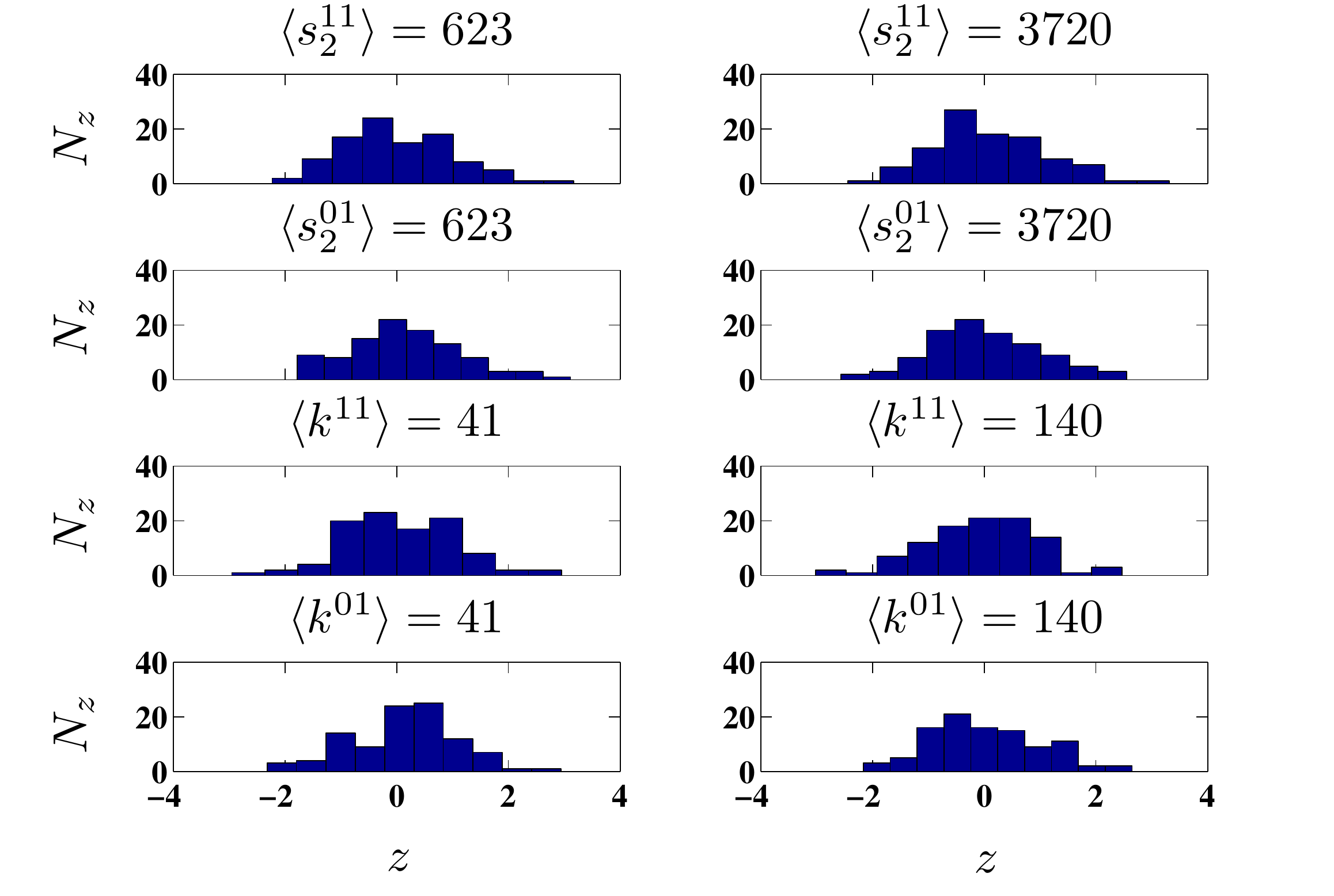}    }
\end{center}
\caption{Distributions of the $z$-scores $\{z_i\}$ related to some fixed values of multistrength and multidegree, in Layer 2 (colorectal cancer layer). In each panel we display the 100 values of $z$ across the sampled instances (gathered in 10 bins), for a chosen node with that assigned value of multistrength or multidegree. Similar results are also found for Layer 1 (normal samples).}
\label{fluctuations}
\end{figure}
\section{Sampling multiplex ensembles with given expected multidegree sequence $\{k_{i}^{\vec{m}}\}$ and given expected multistrength sequence $\{s_{i, \alpha}^{\vec{m}}\}$}
\label{nullmodel}

Here we want to discuss how the theoretical framework described in the previous section can be used to generate weighted multiplex networks sampled from a multiplex network ensemble.
We have chosen to focus specifically on the case of a multiplex network ensemble in which  the given expected multidegree sequence $\{k_{i}^{\vec{m}}\}$ and the given expected multistrength sequence $\{s_{i, \alpha}^{\vec{m}}\}$ are constrained, but the framework we outline here of this case can be easily extended to the other ensembles discussed in this paper.
Given Eqs. (\ref{pstrengthdegreemulti}), (\ref{pistrengthdegreemulti}), the probability $\pi_{ij}(\vec{a}_{ij})$ can be expressed  as a function of the probability $p_{ij}^{\vec{m}}$  of a multilink $\vec{m}$ between node $i$ and node $j$, namely
\small
\begin{equation}
\hspace*{-5mm}\pi_{ij}(\vec{a}_{ij})=p_{ij}^{\vec{m}^{ij}}\prod_{\alpha=1}^{M} \left( \left[e^{-(\lambda_{i,\alpha}^{\vec{m}^{ij}}+\lambda_{j,\alpha}^{\vec{m}^{ij}})} \right]^{a^{\alpha}_{ij}-1} \left[ 1- e^{-(\lambda_{i,\alpha}^{\vec{m}^{ij}}+\lambda_{j,\alpha}^{\vec{m}^{ij}})}\right]
\right)^{m^{ij}_{\alpha}}
\label{pistrengthdegreemultiwithp}
\end{equation}
\normalsize
The productory in Eq. \ref{pistrengthdegreemultiwithp} is the conditional probability of the multiweight $\vec{a}_{ij}$, given the multilink $\vec{m}^{ij}$. The  new expression for $\pi_{ij}(\vec{a}_{ij})$ suggests a way for sampling networks from the distribution given by Eq. $(\ref{probability_wmultiplex})$, with $\pi_{ij}(\vec{a}_{ij})$ given by Eq. $(\ref{pistrengthdegreemultiwithp})$.
In fact for sampling a multiplex network  from this particular ensemble , we draw a multilink $\vec{m}$ with probability  $p_{ij}^{\vec{m}}$ for  each couple of nodes $i$ and $j$. Subsequently, given  a particular multilink, whenever $m_{\alpha}=1$ we draw the additional weight $a^{\alpha}_{ij}-1$ from a geometric distribution with parameter $1- e^{-(\lambda_{i,\alpha}^{\vec{m}^{ij}}+\lambda_{j,\alpha}^{\vec{m}^{ij}})}$ and $a^{\alpha}_{ij}\ge 1$.\\
Following Eqs. (\ref{Lagrangians}) we wrote a Matlab code \cite{SM} that produces the Lagrangian multipliers and calculates the entropy value of the ensemble. The algorithm runs until it finds convergence with precision $10^{-4}$ (this value can be always improved). \\
\section{Comparison between the null model and the biological case study }
Here our aim is to compare the structural properties of our biological case study with the networks with the same multidegree sequence and multistrength sequence generated by sampling the corresponding multiplex network ensemble. 
Starting from our biological duplex network described in section \ref{bio}, at first we calculated the Lagrangian multipliers needed for $\{p_{ij}^{\vec{m}}\}$ and $\{\pi_{ij}(\vec{a}_{ij})\}$, secondly we generated 100 different duplex networks. We checked the average values and fluctuations across our 100 duplexes. In Fig. \ref{average_ensemble_total} we compare the behavior of the average values across the duplexes with the related real values, the assumed fixed average values of the canonical ensemble. We found that the multidegrees and the multistrengths are equal in average to the constrained values showing that the multiplex network framework is able to reproduce well these properties.\\
Nevertheless from sample to sample the individual structural properties of the nodes (their multidegrees and their multistrengths) might fluctuate. In Figure \ref{fluctuations} we investigate the role of the fluctuations by plotting the histogram of the $z$-scores of  values of the multidegrees or of the multistrengths for single nodes, in the layer 2 of the duplex networks, the cancer layer. These distributions are calculated over the 100 multiplex networks sampled by this ensemble.

\section{Conclusions}

In conclusion, in this paper we have characterized the rich interplay between weights and topology of multiplex networks. Multiplex networks describe a large variety of complex systems ranging from social networks to infrastructures and biological networks. Many of these multilayer structures are formed by weighted links, indicating interactions of different intensity. For example, in transportation networks different connections are characterized by a different flow of traffic, in citation and collaboration networks the interactions can be weighted by the number of collaborators or mutual citations, and in biological networks the weights can be given by the strength of chemical bonding or by mutual coexpression measurements. The correlations between weights and topology in these networks can be captured by the multistrength and multi inverse participation ratio. 
As an example, we show that multiplex observables highlihgt significant differences and nontrivial similarities between biological processes in healthy and cancer cells in a gene expression profiling dataset. 

In this paper we provide a framework based on entropy of multiplex networks that can be used  to construct multiplex weighted networks with different level of correlations between the weights and the topology of these structures. Moreover, we have shown how this framework can be applied to generate null models of complex multilayer networks. We believe that such framework can help to develop new methods to shed light on different properties of multiplex networks that cannot be inferred if the single layers were taken separately.

\section{Acknowledgements}
G. M.  acknowledges the kind hospitality of Queen Mary University of London. G. M. and  D. R. acknowledge support by the Italian Ministry of Education and Research through the Flagship (PB05) InterOmics project, the EU MIMOmics (305280) project, and the INFN Gruppo IV Pieces initiative.

\section{Appendix}
\section{Examples of uncorrelated weighted  multiplex ensembles}
\subsection{Multiplex ensembles with given expected total strength in each layer}
As a first example of uncorrelated weighted multiplex, we consider the case in which we  fix the average strength in each layer $\alpha$ to be equal to $S^{\alpha}$. In this case we have $K=M$ constraints in the system, indicated with a label $\alpha=1,2,\ldots,M$. These constraints are given by
\begin{equation}
\sum_{\vec{G}}F_{\alpha}(\vec{G}) P(\vec{G})=\sum_{\vec{G}}\left ( \sum_{i<j} a_{ij}^{\alpha}\right)P(\vec{G})=S^{\alpha}.
\end{equation}
The probability distribution of a multiplex in this ensemble is given by Eq. $(\ref{PC})$ that reads in this case,
\begin{equation}
P(\vec{G})=\frac{1}{Z}\exp \left [ -\sum_{\alpha=1}^M \lambda_{\alpha} \sum_{i<j} a_{ij}^{\alpha}\right],
\end{equation}
where the  partition function $Z$ can be expressed explicitly as
\begin{align}
Z&=\sum_{\vec{G}}\exp \left [ -\sum_{\alpha=1}^M \lambda_{\alpha} \sum_{i<j} a_{ij}^{\alpha}\right]\\
&=\prod_{\alpha=1}^{M}\left[ \left (\frac{1}{1-e^{-\lambda_{\alpha}}}\right )^{\binom{N}{2}}\right]\nonumber.
\end{align} 
The Lagrangian multipliers $\lambda_{\alpha}$ defining the probability of the multiplex $P(\vec{G})$, are fixed by the conditions
\bea
S_{\alpha}&=&-\frac{\partial{\log Z}}{\partial{\lambda_{\alpha}}}=\binom{N}{2}\frac{e^{-\lambda_{\alpha}}}{1-e^{-\lambda_{\alpha}}}.
\label{b5}
\eea
Finally the  average weight $\Avg{a_{ij}^{\alpha}}$ can be evaluated from  Eq. (\ref{av_weight}) and is given by
\bea
\Avg{a_{ij}^{\alpha}}=\frac{S_{\alpha}}{\binom{N}{2}},
\eea
that is equivalent to say $S_{\alpha}=\sum_{i<j}\Avg{a_{ij}^{\alpha}}$.\\
From Eq. (\ref{marginals}) we write the marginal probabilities $\pi(a_{ij}^{\alpha})$ in this specific multiplex ensemble as
\begin{equation}
\label{marginalsS}
\pi_{ij}^{\alpha}(a_{ij}^{\alpha})=e^{-\lambda_{\alpha}a_{ij}^{\alpha}}(1-e^{-\lambda_{\alpha}}).
\end{equation}

Moreover, from Eq. (\ref{probability}) the probability $p_{ij}^{\alpha}$ of having a positive  weight $a_{ij}^{\alpha}>0$ of the link between node $i$ and node $j$ in layer $\alpha$ is independent on the pair of nodes $(i,j)$, i.e. $p_{ij}^{\alpha}=p^{\alpha}$ and is given by 
\begin{equation}
p^{\alpha}=e^{-\lambda_{\alpha}}.
\end{equation}
Finally, the  the probability of a multiplex in this ensemble is given by Eq. (\ref{PCl}) with the marginals $\pi_{ij}^{\alpha}(a_{ij}^{\alpha})$ given by Eq. $(\ref{marginalsS})$. The entropy ${\cal S}$ of this canonical multiplex ensemble is given by 
Eq. $(\ref{entropyS})$. Using the marginals $\pi_{ij}^{\alpha}(a_{ij}^{\alpha})$ given by Eqs.~$(\ref{marginalsS})$ and Eq. $(\ref{b5})$ the entropy can be rearranged as
\bea
\mathcal{S}&=&\sum_{\alpha=1}^M \left [\left (\binom{N}{2}+S_{\alpha} \right)\log \left (\binom{N}{2}+S_{\alpha} \right)\right.\nonumber \\
&&\left.-S_{\alpha}\log S_{\alpha}-\binom{N}{2}\log \binom{N}{2} \right]
\eea
If the number of layers $M$ is finite, applying the Stirling's approximation in the large $N$ limit we get
\begin{equation}
\mathcal{S}=\sum_{\alpha=1}^M \log \left [ \binom{\binom{N}{2}+S^{\alpha}}{\binom{N}{2}} \right].
\end{equation}

\subsection{Multiplex ensembles with given expected strength sequence in each layer}
We consider here the multiplex ensemble in which we  fix the expected strength $s_i^{\alpha}$ of every node $i$, in each layer $\alpha$. We have $K=M \cdot N$ constraints in the system indicated with a label $\alpha=1,2,\ldots,M$. These constraints are given by
\begin{equation}
\sum_{\vec{G}}F_{i,\alpha}(\vec{G}) P(\vec{G})=\sum_{\vec{G}}\left ( \sum_{j \ne i} a_{ij}^{\alpha}\right)P(\vec{G})=s_i^{\alpha}
\end{equation}
The probability of a multiplex $P(\vec{G})$ is given by Eq.(\ref{PC}) that in this case  can be written as
\begin{equation}
P(\vec{G})=\frac{1}{Z}\exp \left [ -\sum_{\alpha=1}^M \sum_i\lambda_{i,\alpha} \sum_{j \ne i} a_{ij}^{\alpha}\right]
\end{equation}
where the  partition function $Z$ can be expressed explicitly as
\bea
Z&=\sum_{\vec{G}}\exp \left [ -\sum_{\alpha=1}^M \sum_i\lambda_{i,\alpha} \sum_{j \ne i} a_{ij}^{\alpha}\right]\nonumber\\
&=\prod_{\alpha=1}^{M}\prod_{i<j}\left[{1-e^{-(\lambda_{i,\alpha}+\lambda_{j,\alpha})}}\right]^{-1},
\eea
and the  Lagrangian multipliers $\lambda_{i,\alpha}$ are fixed by the condition
\begin{equation}
s_i^{\alpha}=-\frac{\partial{\log Z}}{\partial{\lambda_{i,\alpha}}}=\sum_{j \ne i}\frac{e^{-(\lambda_{i,\alpha}+\lambda_{j,\alpha})}}{1-e^{-(\lambda_{i,\alpha}+\lambda_{j,\alpha})}}.
\label{bij}
\end{equation}
The average weight $\Avg{a_{ij}^{\alpha}}$ given by Eq. (\ref{av_weight}) can be calculated explicitly as a function of the Lagrangian multipliers, giving 
\begin{align}
\Avg{a_{ij}^{\alpha}}&=\frac{e^{-(\lambda_{i,\alpha}+\lambda_{j,\alpha})}}{1-e^{-(\lambda_{i,\alpha}+\lambda_{j,\alpha})}},
\end{align}
which implies, together with Eq. $(\ref{bij})$, $s_i^{\alpha}=\sum_{j \ne i}\Avg{a_{ij}^{\alpha}}$.\\
From Eq. (\ref{marginals}) we write the marginal probabilities $\pi_{ij}^{\alpha}(a_{ij}^{\alpha})$ for specific weight $a_{ij}^{\alpha}$ as
\begin{align}
\pi_{ij}^{\alpha}(a_{ij}^{\alpha})&=e^{-(\lambda_{i,\alpha}+\lambda_{j,\alpha})a_{ij}^{\alpha}}(1-e^{-(\lambda_{i,\alpha}+\lambda_{j,\alpha})}),
\label{mij}
\end{align}
i.e. the weight of a link is distributed exponentially, with a mean that depends both on the pair of linked nodes $(i,j)$ and on the layer $\alpha$.
Moreover, from Eq. (\ref{probability}) we can evaluate  the probability $p_{ij}^{\alpha}$ of having a weight different from zero that is given by 
\begin{align}
p_{ij}^{\alpha}&=e^{-(\lambda_{i,\alpha}+\lambda_{j,\alpha})}.
\end{align}
Finally the  the probability of a multiplex in this ensemble is given by Eq. (\ref{PCl}) with the marginals $\pi_{ij}^{\alpha}(a_{ij}^{\alpha})$ given by Eq. $(\ref{mij})$. Therefore the entropy ${\cal S}$ of this canonical multiplex ensemble is given by Eq. $(\ref{entropyS})$
 with the marginals $\pi_{ij}^{\alpha}(a_{ij}^{\alpha})$ given by Eq. $(\ref{mij})$.
\subsection{Multiplex ensembles with given expected strength sequence, given expected  degree sequence and given expected sequences $\{u_{i}^{\alpha}\}$  in each layer}

The last example of uncorrelated multiplex that we will consider is the one in which we  fix the expected strength $s_i^{\alpha}$, the expected degree $k_i^{\alpha}$ and the expected $u_{i}^{\alpha}$ of every node $i$ in each layer $\alpha$. We have $K=M \cdot 3N$ constraints in the system. These constraints are given by
\begin{align}
\sum_{\vec{G}}F_{i,\alpha}(\vec{G}) P(\vec{G})&=\sum_{\vec{G}}\left ( \sum_{j \ne i} a_{ij}^{\alpha}\right)P(\vec{G})=s_i^{\alpha}\nonumber \\
\sum_{\vec{G}}F_{i,\alpha}(\vec{G}) P(\vec{G})&=\sum_{\vec{G}}\left ( \sum_{j \ne i} \theta(a_{ij}^{\alpha})\right)P(\vec{G})=k_i^{\alpha}\nonumber \\
\sum_{\vec{G}}F_{i,\alpha}(\vec{G}) P(\vec{G})&=\sum_{\vec{G}}\left ( \sum_{j \ne i} (a_{ij}^{\alpha})^2\right)P(\vec{G})=u_i^{\alpha}\nonumber \\
\end{align}
with $\alpha=1,2,\ldots, M$.
We introduce the Lagrangian multipliers $\lambda_{i,\alpha}$ for the first set of $N\cdot M$ constraints, the Lagrangian multipliers $\omega_{i,\alpha}$ for the second set of $N\cdot M$ constraints and the Lagrangian multipliers  $z_{i,\alpha}$ for the third set of $N\cdot M$ constraints. Therefore, the probability $P(\vec{G})$ of a multiplex in this ensemble, of general expression given by Eq.~(\ref{PC}), in this specific example is given by  
\begin{equation}
\begin{split}
P(\vec{G})&=\frac{1}{Z}\exp \left[ -\sum_{\alpha=1}^M \sum_i\lambda_{i,\alpha} \sum_{j \ne i} a_{ij}^{\alpha} \right. \\
& \left . -\sum_{\alpha=1}^M \sum_i \omega_{i,\alpha} \sum_{j \ne i} \theta(a_{ij}^{\alpha})-\sum_{\alpha=1}^M \sum_i z_{i,\alpha} \sum_{j \ne i} (a_{ij}^{\alpha})^2 \right ]\nonumber
\end{split}
\end{equation}

If we define as $I_{ij}^{\alpha}$ the series
\begin{equation}
I_{ij}^{\alpha}=\sum_{a_{ij}^{\alpha}=1}^{S^{\alpha}} \exp \left[-(\lambda_{i,\alpha}+\lambda_{j,\alpha})a_{ij}^{\alpha}-(z_{i,\alpha}+z_{j,\alpha})(a_{ij}^{\alpha})^2 \right],
\label{series}
\end{equation}
where $S^{\alpha}=\sum_{i=1}^Ns_i^{\alpha}$. The sum  $I_{ij}^{\alpha}$ is  convergent when $(z_{i,\alpha}+z_{j,\alpha})>0$, the   partition function $Z$ can be expressed as 
\begin{align}
Z&=\prod_{\alpha=1}^{M}\prod_{i<j}\left [ 1+ e^{-(\omega_{i,\alpha}+\omega_{j,\alpha})}I^{\alpha}_{ij}\right]
\end{align}

The Lagrangian multipliers are fixed by the conditions
\begin{align}
-\frac{\partial{\log Z}}{\partial{\lambda_{i,\alpha}}}&=s_i^{\alpha}\nonumber \\
-\frac{\partial{\log Z}}{\partial{\omega_{i,\alpha}}}&=k_i^{\alpha}\nonumber\\
-\frac{\partial{\log Z}}{\partial{z_{i,\alpha}}}&=u_i^{\alpha}
\end{align}
The average weight of the link $(i,j)$ in layer $\alpha$, i.e. $\Avg{a_{ij}^{\alpha}}$, is given by   Eq. (\ref{av_weight}) that in this case reads 
\bea
&&\Avg{a_{ij}^{\alpha}}=\frac{e^{-(\omega_{i,\alpha}+\omega_{j,\alpha})}}{\left [ 1+ e^{-(\omega_{i,\alpha}+\omega_{j,\alpha})}I^{\alpha}_{ij}\right]} \times\nonumber \\
&&\times\left[ \sum_{a_{ij}^{\alpha}=1}^{S^{\alpha}}a_{ij}^{\alpha} \exp \left(-(\lambda_{i,\alpha}+\lambda_{j,\alpha})a_{ij}^{\alpha}-(z_{i,\alpha}+z_{j,\alpha})(a_{ij}^{\alpha})^2 \right)\right]\nonumber
\eea
From Eq. (\ref{marginals}) we write the marginal probabilities $\pi_{ij}^{\alpha}(a_{ij}^{\alpha})$ for this specific ensemble that is given by 
\begin{align}
\pi_{ij}^{\alpha}(a_{ij}^{\alpha})
&=\frac{e^{-(\lambda_{i,\alpha}+\lambda_{j,\alpha})a_{ij}^{\alpha}-(\omega_{i,\alpha}+\omega_{j,\alpha})\theta(a_{ij}^{\alpha})-(z_{i,\alpha}+z_{j,\alpha})(a_{ij}^{\alpha})^2}}{\left [ 1+ e^{-(\omega_{i,\alpha}+\omega_{j,\alpha})}I^{\alpha}_{ij}\right]}
\label{pu}
\end{align}

Moreover, from Eq. (\ref{probability}) the probability $p_{ij}^{\alpha}$ that the link $(i,j)$ in layer $\alpha$ has weight different from zero is given by 
\begin{equation}
p_{ij}^{\alpha}=\frac{e^{-(\omega_{i,\alpha}+\omega_{j,\alpha})}I^{\alpha}_{ij}}{\left [ 1+ e^{-(\omega_{i,\alpha}+\omega_{j,\alpha})}I^{\alpha}_{ij}\right]}
\end{equation}
The probability of a multiplex in this ensemble is given by Eq. (\ref{PCl}) with the marginals $\pi_{ij}^{\alpha}(a_{ij}^{\alpha})$ given by Eq. $(\ref{pu})$ while  the entropy ${\cal S}$ of this canonical multiplex ensemble is given by Eq. $(\ref{entropyS})$
 with the marginals $\pi_{ij}^{\alpha}(a_{ij}^{\alpha})$ given by Eq. $(\ref{pu})$.
\section{Examples of correlated weighted  multiplex ensembles}
\subsection{Multiplex ensembles with given expected total multistrength $S_{\alpha}^{\vec{m}}$}
Here we consider a  correlated weighted multiplex ensemble,  in which we  fix the total  multistrength $\vec{m}$, given by  $S_{\alpha}^{\vec{m}}$ for a layer  $\alpha$ such that  $m_{\alpha}=1$. Since the number of the possible multistrengths $\vec{m}$ in layer $\alpha$ are given by $M\cdot 2^{M-1}$, this gives a number of constraints that is equal to $K=M\cdot2^{M-1}$. 
These constraints are given by
\begin{equation}
\sum_{\vec{G}}F_{\alpha}^{\vec{m}}(\vec{G}) P(\vec{G})=\sum_{\vec{G}}\left ( \sum_{i<j}A_{ij}^{\vec{m}}a_{ij}^{\alpha}\right)P(\vec{G})=S_{\alpha}^{\vec{m}},
\end{equation}
where the multiadjacency matrix element $A_{ij}^{\vec{m}}$ is defined in Eq. $(\ref{multilink})$.
The canonical probability $P(\vec{G})$ of the multiplex in the ensembles is given by the general expression given in Eq. $(\ref{PC})$ that in this case becomes
\begin{equation}
P(\vec{G})=\frac{1}{Z}\exp \left [ -\sum_{\vec{m}\ne\vec{0}}\sum_{\alpha=1}^M \lambda_{\alpha}^{\vec{m}} \sum_{i<j}A_{ij}^{\vec{m}} a_{ij}^{\alpha}\right]
\end{equation}
where the  partition function $Z$ is given by 
\bea
 Z={\cal Z}^{\binom{N}{2}}
 \eea
 where 
 \bea
 {\cal Z}=\sum_{\vec{m}}\prod_{\alpha=1}^{M}\left ( \frac{e^{-\lambda_{\alpha}^{\vec{m}}}}{1-e^{-\lambda_{\alpha}^{\vec{m}}}}\right )^{m_{\alpha}}
\eea
where now, without loss of generality, if $m_{\alpha}=0$ we put $\lambda_{\alpha}^{\vec{m}}=1/2$. We can do this because the probability of a multiplex does not depend on any of these values, and we need to define them only for simplifying the notation.
The Lagrangian multipliers $\lambda_{\alpha}^{\vec{m}}$ with $m_{\alpha}=1$, are fixed by the conditions
\begin{equation}
-\frac{\partial{\log Z}}{\partial{\lambda_{\alpha}^{\vec{m}}}}=S_{\alpha}^{\vec{m}},
\end{equation}
which yields 
\bea
S_{\alpha}^{\vec{m}}=\binom{N}{2}\frac{1}{\cal Z} \left ( \frac{1}{1-e^{-\lambda_{\alpha}^{\vec{m}}}}\right )\prod_{\beta=1}^{M}\left ( \frac{e^{-\lambda_{\beta}^{\vec{m}}}}{1-e^{-\lambda_{\beta}^{\vec{m}}}}\right )^{m_{\beta}}.
\label{lag0}
\eea

The probability of a multiplex $P(\vec{G})$ follows Eq. \ref{probability_wmultiplex} with
\begin{equation}
\pi_{ij}(\vec{a}_{ij})=\frac{ e^{-\sum_{\alpha=1,M}\lambda_{\alpha}^{\vec{m}^{ij}}a^{\alpha}_{ij}}}{\mathcal{Z}},
\label{pi_prob}
\end{equation}
Further on we can compute the average weight of the link $ij$ on the multilink $\vec{m}$, in the layer $\alpha$
\begin{equation}
\Avg{a_{ij}^{\alpha}A_{ij}^{\vec{m}}}=\sum_{\vec{G}}a_{ij}^{\alpha}A_{ij}^{\vec{m}}P(\vec{G})=\sum_{\vec{a}_{ij}}a_{ij}^{\alpha}A_{ij}^{\vec{m}}\pi(\vec{a}_{ij}).
\end{equation}
Using Eq. $(\ref{pi_prob})$ for the explicit expression of $\pi(\vec{a}_{ij})$ and comparing the results with Eq. $(\ref{lag0})$
it is easy to show that 
\bea
\Avg{a_{ij}^{\alpha}A_{ij}^{\vec{m}}}=\frac{S_{\alpha}^{\vec{m}}}{\binom{N}{2}}.
\eea
The probability of a multilink $\vec{m}$ between node $i$ and node $j$, $p_{ij}^{\vec{m}}=\Avg{A_{ij}^{\vec{m}}}$ in this ensemble is independent on the pair of nodes  $(i,j)$. Therefore we have $p_{ij}^{\vec{m}}=p^{\vec{m}}$ with 
\begin{equation}
p^{\vec{m}}=\frac{\prod_{\alpha=1}^{M}\left ( \frac{e^{-\lambda_{\alpha}^{\vec{m}}}}{1-e^{-\lambda_{\alpha}^{\vec{m}}}}\right )^{m_{\alpha}}}{\mathcal{Z}},
\end{equation}
Finally, the entropy ${\cal S}$ of this ensemble follows Eq. $(\ref{entropy2})$.
\subsection{Multiplex ensembles with given expected $\nu$-total strength $S^{\nu}_{\alpha}$}
In presence of many layers $M$ we can consider as constraints the average  $\nu$-total strength $S^{\nu}_{\alpha}$ with $\nu=1,2,\ldots, M$.
With respect to the previous case, now the number of constraints is sensibly reduced and is given by $M^2$ constraints  
\begin{equation}
\sum_{\vec{G}}F_{\alpha}^{\nu}(\vec{G}) P(\vec{G})=\sum_{\vec{G}}\left ( \sum_{i<j}a_{ij}^{\alpha}A_{ij}^{\nu}\right)P(\vec{G})=S_{\alpha}^{\nu}.
\end{equation}
The probability $P(\vec{G})$ of the multiplex network, is therefore given in terms of $M^2$ Lagrangian multipliers $\lambda_{\alpha}^{\nu}$, i.e. 
\begin{equation}
P(\vec{G})=\frac{1}{Z} \exp\left[- \sum_{\nu=1}^{M}\sum_{\alpha=1}^M \lambda_{\alpha}^{\nu}\sum_{i<j}A_{ij}^{\nu} a_{ij}^{\alpha}\right]
\label{constraintsnu}
\end{equation}
where the  partition function $Z$ is given by $Z={\mathcal Z}^{\binom{N}{2}}$ with 
\begin{equation}
\mathcal{Z}=\sum_{\nu=0}^{M}\sum_{\vec{m}|\nu(\vec{m})=\nu}\prod_{\alpha=1}^{M}\left ( \frac{e^{-\lambda_{\alpha}^{\nu} }}{1-e^{-\lambda_{\alpha}^{\nu}}}\right )^{m_{\alpha}}.
\end{equation}
The Lagrangian multipliers $\lambda_{\alpha}^{\nu}$ are fixed  fixed by the constraints Eq.$(\ref{constraintsnu})$ that can be also expressed as 
\begin{equation}
-\frac{\partial{\log Z}}{\partial{\lambda_{\alpha}^{\nu}}}=S_{\alpha}^{\nu}.
\end{equation}
The probability $P(\vec{G})$ of the multiplex network
is given by Eq.~$(\ref{probability_wmultiplex})$
and the entropy of the ensemble takes the simple expression given by Eq.~ $(\ref{entropy2})$ where $\pi_{ij}(\vec{a}_{ij})$ is given by 
\begin{equation}
\pi_{ij}(\vec{a}_{ij})=\frac{ e^{-\sum_{\alpha=1}^M\lambda_{\alpha}^{\nu^{ij}}a^{\alpha}_{ij}}}{\mathcal{Z}}.
\label{pi_probnu}
\end{equation}

Finally the  probability $p^{\nu}$ of a $\nu$-multilink  between any two nodes of the multiplex network is given by 
\begin{equation}
p^{\nu}=\frac{1}{\mathcal{Z}}\prod_{\alpha=1}^{M}\left ( \frac{e^{-\lambda_{\alpha}^{\nu}}}{1-e^{-\lambda_{\alpha}^{\nu}}}\right )^{m_{\alpha}},
\end{equation}
while we have that the average weight of a $\nu$ mutlilink is given by 
\bea
\Avg{a_{ij}^{\alpha}A_{ij}^{\nu}}&=&\frac{S_{\alpha}^{\nu}}{\binom{N}{2}}=\frac{1}{\mathcal{Z}} \left ( \frac{1}{1-e^{-\lambda_{\alpha}^{\nu}}}\right )\times \nonumber \nonumber \\
&&\times\sum_{\vec{m}|\nu(\vec{m})=\nu}m_{\alpha}\prod_{\beta=1}^{M}\left ( \frac{e^{-\lambda_{\beta}^{\nu}}}{1-e^{-\lambda_{\beta}^{\nu}}}\right )^{m_{\beta}}.
\eea

\subsection{Multiplex ensembles with given expected multistrength sequence $\{s_{i, \alpha}^{\vec{m}}\}$}
Here we consider another level  of coarse-graining for the multiplex network and we study  correlated weighted multiplex in which we fix the average strength sequence $s_{i, \alpha}^{\vec{m}}$ for each node $i$, in each layer $\alpha$ such that $m_{\alpha}=1$, for a given multilink $\vec{m}$. Following the previous line of reasoning, we can express properly just $N \cdot M \cdot 2^{M-1}$ constraints.\\
These constraints are given by
\begin{equation}
\sum_{\vec{G}}F_{i, \alpha}^{\vec{m}}(\vec{G}) P(\vec{G})=\sum_{\vec{G}}\left ( \sum_{j\ne i}A_{ij}^{\vec{m}}a_{ij}^{\alpha}\right)P(\vec{G})=s_{i, \alpha}^{\vec{m}},
\end{equation}
with $i=1,\ldots, N$, $\vec{m}=(m_1,m_2,\ldots,m_{\beta}, \ldots, m_M)$ with $m_{\beta}=0,1$ and finally $\alpha=1,\ldots, M$ with the condition $m_{\alpha}=1$.
The canonical probability $P(\vec{G})$ of the multiplex in the ensemble is
\bea
P(\vec{G})&=&\frac{1}{Z}\exp \left [ -\sum_{\vec{m}\ne\vec{0}}\sum_{\alpha=1}^M \sum_i \lambda_{i,\alpha}^{\vec{m}} \sum_{j \ne i}A_{ij}^{\vec{m}} a_{ij}^{\alpha}\right]\nonumber\\
&&\hspace*{-17mm}=\frac{1}{Z}\prod_{i<j}\exp \left [ -\sum_{\vec{m}\ne\vec{0}}\sum_{\alpha=1}^M (\lambda_{i,\alpha}^{\vec{m}}+\lambda_{j,\alpha}^{\vec{m}})A_{ij}^{\vec{m}} a_{ij}^{\alpha}\right],
\label{PGm2}
\eea
where the partition function $Z$ can be expressed explicitly as
\bea
Z&=&\prod_{i<j}{\cal Z}_{ij}
\eea
where 
\bea
{\cal Z}_{ij}=\sum_{\vec{m}}\prod_{\alpha=1}^{M}\left ( \frac{e^{-(\lambda_{i,\alpha}^{\vec{m}} +\lambda_{j,\alpha}^{\vec{m}}  )}}{1-e^{-(\lambda_{i,\alpha}^{\vec{m}} +\lambda_{j,\alpha}^{\vec{m}}  )}}\right )^{m_{\alpha}},
\eea
where now, without loss of generality, if $m_{\alpha}=0$ we put $\lambda_{\alpha}^{\vec{m}}=1/2$. We can do this because the probability of a multiplex and the partition function do not depend on any of these values, and we need to define them only for simplifying the notation.
The Lagrangian multipliers $\lambda_{i,\alpha}^{\vec{m}}$, with $\alpha$ such that $m_{\alpha}=1$, are fixed by the conditions
\begin{equation}
-\frac{\partial{\log Z}}{\partial{\lambda_{i,\alpha}^{\vec{m}}}}=s_{i,\alpha}^{\vec{m}}=\sum_{j \ne i}\Avg{a_{ij}^{\alpha}A_{ij}^{\vec{m}}},
\end{equation}
where $\Avg{a_{ij}^{\alpha}A_{ij}^{\vec{m}}}$ is the average weight of the link between node$i$ and node $j$ on the multilink $\vec{m}$, in the layer $\alpha$. This quantity  can be computed as
\begin{equation}
\hspace*{-7mm}\Avg{a_{ij}^{ \alpha}A_{ij}^{\vec{m}}}=\frac{1}{\mathcal{Z}_{ij}} \left(\frac{1}{1-e^{-(\lambda_{i,\alpha}^{\vec{m}}+\lambda_{j,\alpha}^{\vec{m}})}}\right)\prod_{\beta=1}^{M}\left ( \frac{e^{-(\lambda_{i,\beta}^{\vec{m}}+\lambda_{j,\beta}^{\vec{m}})}}{1-e^{-(\lambda_{i,\beta}^{\vec{m}}+\lambda_{j,\beta}^{\vec{m}})}}\right )^{m_{\beta}}.\nonumber
\label{lag}
\end{equation}
We can calculate the probability of a vector $\vec{a}_{ij}=(a_{ij}^1,a_{ij}^2\ldots, a_{ij}^M)$ characterizing the weights of the links between node $i$ and node $j$ in all the layers, getting
\begin{equation}
\pi_{ij}(\vec{a}_{ij})=\frac{1}{\mathcal{Z}_{ij}} { e^{-\sum_{\alpha=1,M}(\lambda_{i,\alpha}^{\vec{m}^{ij}}+\lambda_{j,\alpha}^{\vec{m}^{ij}})a^{\alpha}_{ij}}}.
\label{piijvm2}
\end{equation}
These probabilities satisfy the normalization condition given by Eq. $(\ref{piijnorm})$.
The probability $p_{ij}^{\vec{m}}$ of a multilink $\vec{m}$ between the node $i$ and the node $j$ is given by
\begin{equation}
p_{ij}^{\vec{m}}=\Avg{A_{ij}^{\vec{m}}}=\frac{1}{\mathcal{Z}_{ij}}{\prod_{\alpha=1}^{M}\left ( \frac{e^{-(\lambda_{i,\alpha}^{\vec{m}} +\lambda_{j,\alpha}^{\vec{m}}  )}}{1-e^{-(\lambda_{i,\alpha}^{\vec{m}} +\lambda_{j,\alpha}^{\vec{m}}  )}}\right )^{m_{\alpha}}},
\label{pijvm2}
\end{equation}
where these probabilities satisfy the normalization condition given by Eq. $(\ref{pijnorm})$ and are related to the probabilities $\pi_{ij}^{\vec{m}}(\vec{a}_{ij})$ given by Eq. $(\ref{piijvm2})$, by Eq. $(\ref{pijpiij1})$.

Probability $P(\vec{G})$ and entropy $\mathcal{S}$ follow Eqs. (\ref{probability_wmultiplex}), (\ref{entropy2}) respectively.
\subsection{Multiplex ensembles with given expected $\nu$-multistrength sequence $\{s_{i,\alpha}^{\nu} \}$ }

In the case in which one wants to describe multiplex networks with many layers $M$, one can consider to fix the average $\nu$-multistrength sequence $\{s_{i, \alpha}^{\nu}\}$
 with  $i=1,2\ldots, N$ and $\nu=1,2,\ldots, M$.
Therefore, the number of  constraints of the previous example is reduced to just  $N \cdot M^2$ soft constraints given by 
\begin{equation}
\sum_{\vec{G}}F_{i, \alpha}^{\nu}(\vec{G}) P(\vec{G})=\sum_{\vec{G}}\left ( \sum_{j\ne i}a_{ij}^{\alpha}A_{ij}^{\nu}\right)P(\vec{G})=s_{i, \alpha}^{\nu}.
\label{conditionsnu2}
\end{equation}
In this case, the probability $P(\vec{G})$ of a multiplex network  $\vec{G}$ in this ensemble  is expressed in terms of the $N \times M^2$ Lagrangian multipliers $\lambda_{i,\alpha}^{\nu}$ and is given by 
\begin{equation}
P(\vec{G})= \frac{1}{Z}\exp\left[-\sum_{i<j} \sum_{\nu=1}^{M}\sum_{\alpha=1}^M (\lambda_{i,\alpha}^{\nu}+\lambda_{j,\alpha}^{\nu})A_{ij}^{\nu} a_{ij}^{\alpha}\right],\nonumber
\end{equation}
where the  partition function $Z$ can be expressed as
\begin{align}
Z&=\prod_{i<j}\mathcal{Z}_{ij}
\end{align}
with   
\begin{equation}
\mathcal{Z}_{ij}=\sum_{\nu=0}^{M}\sum_{\vec{m}|\nu(\vec{m})=\nu}\prod_{\alpha=1}^{M}\left ( \frac{e^{-(\lambda_{i,\alpha}^{\nu} +\lambda_{j,\alpha}^{\nu}  )}}{1-e^{-(\lambda_{i,\alpha}^{\nu} +\lambda_{j,\alpha}^{\nu}  )}}\right )^{m_{\alpha}}.
\end{equation}

The Lagrangian multipliers are fixed by the conditions in Eq. ~$(\ref{conditionsnu2})$, or equivalently by 
\begin{equation}
-\frac{\partial{\log Z}}{\partial{\lambda_{i,\alpha}^{\nu}}}=s_{i,\alpha}^{\nu}=\sum_{j \ne i}\Avg{a_{ij}^{\alpha}A_{ij}^{\nu}}.
\end{equation}
Therefore the probability $P(\vec{G})$ of a multiplex network $\vec{G}$ in this ensemble, 
is given by Eq.~$(\ref{probability_wmultiplex})$
and the entropy of the ensemble takes the simple expression given by Eq.~ $(\ref{entropy2})$ where $\pi_{ij}(\vec{a}_{ij})$ is given by 
\begin{equation}
\pi_{ij}(\vec{a}_{ij})=\frac{1}{\mathcal{Z}_{ij}} { e^{-\sum_{\alpha=1,M}(\lambda_{i,\alpha}^{\nu^{ij}}+\lambda_{j,\alpha}^{\nu^{ij}})a^{\alpha}_{ij}}}.
\end{equation}
Finally, the probability $p_{ij}^{\nu}$ that the node $i$ and the node $j$ are linked by a $\nu$-multilink is given by 
\bea
p_{ij}^{\nu}&=&\frac{1}{\mathcal{Z}_{ij}}\sum_{\vec{m}|\nu(\vec{m})=\nu}\prod_{\alpha=1}^{M}\left ( \frac{e^{-(\lambda_{i,\alpha}^{\nu} +\lambda_{j,\alpha}^{\nu}  )}}{1-e^{-(\lambda_{i,\alpha}^{\nu} +\lambda_{j,\alpha}^{\nu}  )}}\right )^{m_{\alpha}},
\eea
while the average weight of the link $a_{ij}^{\alpha}$ belonging to a $\nu$-multilink is given by 
\bea
\Avg{a_{ij}^{\alpha}A_{ij}^{\nu}}&=&\frac{1}{\mathcal{Z}_{ij}} \left ( \frac{1}{1-e^{-(\lambda_{i,\alpha}^{\nu}+\lambda_{j,\alpha}^{\nu})}}\right )\times \nonumber \nonumber \\
&&\hspace*{-10mm}\times\sum_{\vec{m}|\nu(\vec{m})=\nu}m_{\alpha}\prod_{\beta=1}^{M}\left ( \frac{e^{-(\lambda_{i,\beta}^{\nu}+\lambda_{j,\beta}^{\nu})}}{1-e^{-(\lambda_{i,\beta}^{\nu}+\lambda_{j,\beta}^{\nu})}}\right )^{m_{\beta}}.
\eea
\subsection{Multiplex ensembles with given expected multidegree sequence $\{k_{i}^{\vec{m}}\}$, given expected multistrength sequence $\{s_{i, \alpha}^{\vec{m}}\}$ and given expected sequence $\{u_{i, \alpha}^{\vec{m}}\}$}
As a fourth case of correlated weighted multiplex ensemble, we consider the case in which we fix the average multidegree $k_{i}^{\vec{m}}$ of node $i$, for each node $i=1,\ldots,N$, for $\vec{m}\neq\vec{0}$. Moreover, for each node $i$ in layer $\alpha$ we impose the  average multistrength   $s_{i, \alpha}^{\vec{m}}$ and the second moment of the weights incident to it and belonging to a multilink $\vec{m}$, i.e. $u^{\vec{m}}_{i,\alpha}$. The number of independent constraints is therefore $K=(2^{M}-1) \cdot N + 2^{M}\cdot M \cdot N$.\\
In particular, the constraints we are imposing are the following, 
\begin{align}
\sum_{\vec{G}}F_{i, \alpha}^{\vec{m}}(\vec{G}) P(\vec{G})&=\sum_{\vec{G}}\left ( \sum_{j\ne i}A_{ij}^{\vec{m}}a_{ij}^{\alpha}\right)P(\vec{G})=s_{i, \alpha}^{\vec{m}}\nonumber\\
\sum_{\vec{G}}F_{i}^{\vec{m}}(\vec{G}) P(\vec{G})&=\sum_{\vec{G}}\left ( \sum_{j\ne i}A_{ij}^{\vec{m}}\right)P(\vec{G})=k_{i}^{\vec{m}}\nonumber\\
\sum_{\vec{G}}F_{i, \alpha}^{\vec{m}}(\vec{G}) P(\vec{G})&=\sum_{\vec{G}}\left ( \sum_{j\ne i}(A_{ij}^{\vec{m}}a_{ij}^{\alpha})^2\right)P(\vec{G})=u^{\vec{m}}_{i,\alpha}
\end{align}
The canonical probability $P(\vec{G})$ of the multiplex in the ensembles is
\bea
P(\vec{G})&=&\frac{1}{Z}\exp \left [ -\sum_{i<j}\sum_{\vec{m}\neq \vec{0}}(\omega_i^{\vec{m}}+\omega_j^{\vec{m}})A_{ij}^{\vec{m}}\right]\times \\
&\times & \exp\left[-\sum_{i<j} \sum_{\vec{m}\neq\vec{0}}\sum_{\alpha=1}^M (\lambda_{i,\alpha}^{\vec{m}}+\lambda_{j,\alpha}^{\vec{m}})A_{ij}^{\vec{m}} a_{ij}^{\alpha}\right] \nonumber\\
&\times & \exp\left[-\sum_{i<j} \sum_{\vec{m}\neq\vec{0}}\sum_{\alpha=1}^M (z_{i,\alpha}^{\vec{m}}+z_{j,\alpha}^{\vec{m}})A_{ij}^{\vec{m}} (a_{ij}^{\alpha})^2\right]\nonumber
\eea
The partition function $Z$ can be expressed explicitly as
\bea
Z&=&\prod_{i<j}{\mathcal {Z}}_{ij}\nonumber \\
&=&\prod_{i<j}\left(1+\sum_{\vec{m}\ne \vec{0}}e^{-(\omega_i^{\vec{m}}+\omega_j^{\vec{m}})}\prod_{\alpha=1}^{M}\left ( I_{ij}^{\vec{m}, \alpha}\right )^{m_{\alpha}}\right)
\eea
where  $I_{ij}^{\vec{m}, \alpha}$ is given by

\begin{equation}
I_{ij}^{\vec{m}, \alpha}=\sum_{a_{ij}^{\alpha}=1}^{S^{\vec{m},\alpha}}\exp \left[-(\lambda_{i,\alpha}^{\vec{m}}+\lambda_{j, \alpha}^{\vec{m}})a_{ij}^{\alpha}-(z_{i, \alpha}^{\vec{m}}+z_{j, \alpha}^{\vec{m}})(a_{ij}^{\alpha})^2 \right]\nonumber,
\end{equation}
where $S^{\vec{m},\alpha}=\sum_{i=1}^N s_{i,\alpha}^{\vec{m}}$.
The Lagrangian multipliers are fixed by the conditions
\begin{align}
-\frac{\partial{\log Z}}{\partial{\lambda_{i,\alpha}^{\vec{m}}}}&=s_{i,\alpha}^{\vec{m}}=\sum_{j \ne i}\Avg{a_{ij}^{\alpha}A_{ij}^{\vec{m}}},\nonumber\\
-\frac{\partial{\log Z}}{\partial{\omega_{i}^{\vec{m}}}}&=k_{i}^{\vec{m}}=\sum_{j \ne i}\Avg{A_{ij}^{\vec{m}}},\nonumber\\
-\frac{\partial{\log Z}}{\partial{z_{i,\alpha}^{\vec{m}}}}&= u_{i,\alpha}^{\vec{m}}=\sum_{j \ne i}\Avg{(a_{ij}^{\alpha})^2A_{ij}^{\vec{m}}}
\end{align}

The average weight $\Avg{a_{ij}^{\alpha}A_{ij}^{\vec{m}}}$ of the multilink $\vec{m}$ between nodes $i$ and $j$  in the layer $\alpha$ and  the probability $p_{ij}^{\vec{m}}$  of a multilink $\vec{m}$ between node $i$ and node $j$ are given respectively by 
\bea
\Avg{a_{ij}^{\alpha}A_{ij}^{\vec{m}}}&=&-\frac{e^{-(\omega_{i}^{\vec{m}}+\omega_{j}^{\vec{m}})}}{\mathcal{Z}_{ij}} \left ( \frac{\partial I_{ij}^{\vec{m}, \alpha}}{\partial(\lambda_{i,\alpha}^{\vec{m}}+\lambda_{j,\alpha}^{\vec{m}})}\right )\times \nonumber \\
&&\times\prod_{\beta \ne \alpha}^{M}\left ( I_{ij}^{\vec{m}, \beta}\right )^{m_{\beta}}\nonumber\\
\hspace*{-3mm}p_{ij}^{\vec{m}}&\hspace*{-3mm}=&\hspace*{-3mm}\frac{e^{-(\omega_i^{\vec{m}}+\omega_j^{\vec{m}})}}{\mathcal{Z}_{ij}}\prod_{\alpha=1}^{M}\left ( I_{ij}^{\vec{m}, \alpha}\right )^{m_{\alpha}}
\eea

The probability of a specific multiweight $\vec{a}_{ij}$ in the between the nodes  $(i,j)$ is
\bea
\pi_{ij}(\vec{a}_{ij})&=&\frac{e^{-(\omega_i^{\vec{m}^{ij}}+\omega_j^{\vec{m}^{ij}})}}{\mathcal{Z}_{ij}} e^{-\sum_{\alpha=1,M}(\lambda_{i,\alpha}^{\vec{m}^{ij}}+\lambda_{j,\alpha}^{\vec{m}^{ij}})a^{\alpha}_{ij}}\times \nonumber \\
&\times & e^{-\sum_{\alpha=1,M}(z_{i,\alpha}^{\vec{m}^{ij}}+z_{j,\alpha}^{\vec{m}^{ij}})(a_{ij}^{\alpha})^2}
\eea
As previously, probability $P(\vec{G})$ and entropy $\mathcal{S}$ follow Eqs. (\ref{probability_wmultiplex}), (\ref{entropy2}) respectively.
\subsection{Multiplex ensembles with given expected $\nu$-multidegree sequence $\{k_{i}^{\nu} \}$, given expected $\nu$-multistrength sequence $\{s_{i,\alpha}^{\nu} \}$ and given expected sequence $\{u_{i, \alpha}^{\nu}\}$ }
Finally we consider the case in which we fix the given expected $\nu$-multidegree sequence $k_{i}^{\nu}$ of node $i$, for each node $i=1,\ldots,N$, for $\nu=1, ..., M$. In addition, for each node $i$ in layer $\alpha$ we fix the  average $\nu$-multistrength sequence   $s_{i, \alpha}^{\nu}$ and the second moment of the weights incident to it for each $\nu$-multilink, i.e. $u^{\nu}_{i,\alpha}$. The $N\cdot M \cdot (2M+1)$ constraints are given by
\begin{align}
\sum_{\vec{G}}F_{i, \alpha}^{\nu}(\vec{G}) P(\vec{G})&=\sum_{\vec{G}}\left ( \sum_{j\ne i}a_{ij}^{\alpha}A_{ij}^{\nu}\right)P(\vec{G})=s_{i, \alpha}^{\nu}\nonumber\\
\sum_{\vec{G}}F_{i}^{\nu}(\vec{G}) P(\vec{G})&=\sum_{\vec{G}}\left ( \sum_{j\ne i}A_{ij}^{\nu}\right)P(\vec{G})=k_{i}^{\nu}\nonumber\\
\sum_{\vec{G}}F_{i, \alpha}^{\nu}(\vec{G}) P(\vec{G})&=\sum_{\vec{G}}\left ( \sum_{j\ne i}(a_{ij}^{\alpha}A_{ij}^{\nu})^2\right)P(\vec{G})=u_{i, \alpha}^{\nu}
\label{conditionsnu4}
\end{align}
with $i=1,2,\ldots, N$, $\alpha=1,2,\ldots M$ and $\nu=1,2,\ldots, M$.
The canonical probability $P(\vec{G})$ of the multiplex in this ensemble can be expressed in terms of the Lagrangian multipliers $\lambda_{j,\alpha}^{\nu}$, $ \omega_i^{\nu}$ and $z_{j,\alpha}^{\nu}$, i.e. 
\bea
P(\vec{G})&=&\frac{1}{Z}\exp \left [ -\sum_{i<j}\sum_{\nu=1}^{M}(\omega_i^{\nu}+\omega_j^{\nu})A_{ij}^{\nu}\right]\times \\
&\times & \exp\left[-\sum_{i<j} \sum_{\nu=1}^{M}\sum_{\alpha=1}^M (\lambda_{i,\alpha}^{\nu}+\lambda_{j,\alpha}^{\nu})A_{ij}^{\nu} a_{ij}^{\alpha}\right]\nonumber\\
&\times & \exp\left[-\sum_{i<j} \sum_{\nu=1}^{M}\sum_{\alpha=1}^M (z_{i,\alpha}^{\nu}+z_{j,\alpha}^{\nu})A_{ij}^{\nu} (a_{ij}^{\alpha})^2\right]\nonumber
\eea
where the  partition function $Z$ is given by 
\begin{align}
Z&=\prod_{i<j}\mathcal{Z}_{ij},
\end{align}
with   
\begin{equation}
\mathcal{Z}_{ij}=1+\sum_{\nu=1}^{M}e^{-(\omega_i^{\nu}+\omega_j^{\nu})}\sum_{\vec{m}|\nu(\vec{m})=\nu}\prod_{\alpha=1}^{M} (I_{ij}^{\nu, \alpha} )^{m_{\alpha}}
\end{equation}
where  $I_{ij}^{\nu, \alpha}$ is given by
\begin{equation}
I_{ij}^{\nu, \alpha}=\sum_{a_{ij}^{\alpha}=1}^{S^{\nu,\alpha}}\exp \left[-(\lambda_{i,\alpha}^{\nu}+\lambda_{j, \alpha}^{\nu})a_{ij}^{\alpha}-(z_{i, \alpha}^{\nu}+z_{j, \alpha}^{\nu})(a_{ij}^{\alpha})^2 \right]\nonumber,
\end{equation}
where $S^{\nu,\alpha}=\sum_{i=1}^N s_{i,\alpha}^{\nu}$

The Lagrangian multipliers are fixed by the conditions Eq.~$(\ref{conditionsnu4})$ that can be also written in terms of the partial derivatives of the partition function as 
\bea
-\frac{\partial{\log Z}}{\partial{\lambda_{i,\alpha}^{\nu}}}&=&s_{i,\alpha}^{\nu}=\sum_{j \ne i}\Avg{a_{ij}^{\alpha}A_{ij}^{\nu}},\nonumber\\
-\frac{\partial{\log Z}}{\partial{\omega_{i}^{\nu}}}&=&k_{i}^{\nu}=\sum_{j \ne i}\Avg{A_{ij}^{\nu}}\nonumber\\
-\frac{\partial{\log Z}}{\partial{z_{i,\alpha}^{\nu}}}&=&u_{i, \alpha}^{\nu}=\sum_{j \ne i}\Avg{(a_{ij}^{\alpha})^2A_{ij}^{\nu}}.
\eea
The probability $P(\vec{G})$ of a multiplex network $\vec{G}$ and the consequent entropy of the ensemble  are given by Eq.~$(\ref{probability_wmultiplex})$ and Eq.~ $(\ref{entropy2})$ with $\pi_{ij}(\vec{a}_{ij})$ given by 
\bea
\pi_{ij}(\vec{a}_{ij})&=&\frac{e^{-(\omega_i^{\nu^{ij}}+\omega_j^{\nu^{ij}})}}{\mathcal{Z}_{ij}} { e^{-\sum_{\alpha=1,M}(\lambda_{i,\alpha}^{\nu^{ij}}+\lambda_{j,\alpha}^{\nu^{ij}})a^{\alpha}_{ij}}}\times \nonumber \\
&\times & e^{-\sum_{\alpha=1,M}(z_{i,\alpha}^{\nu^{ij}}+z_{j,\alpha}^{\nu^{ij}})(a_{ij}^{\alpha})^2}
\eea

The probability $p_{ij}^{\nu}$ that the node $i$ and the node $j$ are linked by a $\nu$-multilink is given by 
\bea
p_{ij}^{\nu}&=&\frac{e^{-(\omega_i^{\nu}+\omega_j^{\nu})}}{\mathcal{Z}_{ij}}\sum_{\vec{m}|\nu(\vec{m})=\nu}\prod_{\alpha=1}^{M} (I_{ij}^{\nu, \alpha})^{m_{\alpha}}
\eea
Finally, the average weight of the link $a_{ij}^{\alpha}$ belonging to a $\nu$-multilink is given by 
\bea
\Avg{a_{ij}^{\alpha}A_{ij}^{\nu}}&=&-\frac{e^{-(\omega_{i}^{\nu}+\omega_{j}^{\nu})}}{\mathcal{Z}_{ij}} \left ( \frac{\partial I_{ij}^{\nu, \alpha}}{\partial(\lambda_{i,\alpha}^{\nu}+\lambda_{j,\alpha}^{\nu})}\right )\times \nonumber \nonumber \\
&\times &\sum_{\vec{m}|\nu(\vec{m})=\nu}m_{\alpha}\prod_{\beta \ne \alpha}^{M}(I_{ij}^{\nu, \beta})^{m_{\beta}}
\eea


\begin{thebibliography}{99}

\bibitem{PhysReport}
 S. Boccaletti, G. Bianconi, R. Criado, C.I. del Genio, J. G\'omez-Garde\~nes, M. Romance, I. Sendi\~na-Nadal, Z. Wang, M. Zanin, Physics Reports {\bf 544}, 1 (2014).

\bibitem{Kivela}
M. Kivel\"a et al Journal of Complex Networks {\bf 2}, 203 (2014).

\bibitem{Thurner}
M. Szell, R. Lambiotte, S. Thurner, PNAS, {\bf 107}, 13636 (2010).

\bibitem{Menichetti}
G. Menichetti, D. Remondini, P. Panzarasa, R. J. Mondrag\'on G. Bianconi  PLoS ONE {\bf 9}, e97857 (2014).

\bibitem{Boccaletti}
A. Cardillo, J. G\'omez-Garde\~nes, M. Zanin, M. Romance, D. Papo, F. del Pozo  S. Boccaletti, Sci. Rep. {\bf 3}, 1344 (2013).



\bibitem{Vito}
 V. Nicosia and V. Latora, arXiv:1403.1546 (2014).

\bibitem{Kurths} J. Donges, H. Schultz, N. Marwan, Y. Zou and
  J. Kurths,  {The European Physical Journal B} \textbf{84},

\bibitem{Bullmore} E. Bullmore and O. Sporns,  {Nat Rev
  Neurosci} \textbf{10}, 186-198 (2009).

\bibitem{GDneuro} G. Castellani, N. Intrator and D. Remondini, {Frontiers in Genetics} \textbf{5:253}, 10.3389/fgene.2014.00253 (2014)

\bibitem{Netonet}
G. Bianconi and S.N. Dorogovtsev  Phys. Rev. E 89, 062814 (2014).

\bibitem{Havlin3}
J. Gao, S.V. Buldyrev, H.E. Stanley, S. Havlin, Nature Physics{\bf 8}, 40 (2012).
   
\bibitem{Mucha}
 P. J. Mucha, T. Richardson, K. Macon, M. A Porter, J.-P. Onnela,
Science, {\bf 328},876 (2010).   

\bibitem{Battiston}
 F. Battiston, V. Nicosia and V. Latora,  Phys. Rev. E {\bf 89}, 032804 (2014).

\bibitem{Goh}
Byungjoon Min, Su Do Yi, Kyu-Min Lee, and K.-I. Goh,  Phys. Rev. E 89, 042811  (2014).

\bibitem{PRL}
V. Nicosia, G. Bianconi, V. Latora, and M. Barthelemy, Phys. Rev. Lett. 111,058701 (2013).

\bibitem{PRE}
G. Bianconi, Phys. Rev. E  {\bf 87}, 062806 (2013).

\bibitem{Nonlinear}
V. Nicosia, G. Bianconi, V. Latora, and M. Barthelemy,arXiv: arXiv:1312.3683 (2013).

\bibitem{Kertesz}
M. Yohsuke, J. T\"or\"ok, H.-H. Jo, K. Kaski,  J. Kert\'esz, Physical Review E {\bf 90},  052810 (2014).

\bibitem{Spatial}
A. Halu, S. Mukherjee, G. Bianconi, Phys. Rev. E 89, 012806 (2014) 

\bibitem{Havlin1}
S.V. Buldyrev, R. Parshani, G. Paul, H.E. Stanley, S. Havlin, Nature 
{\bf  464}, 1025 (2010) 

\bibitem{Diffusion} S. G\'omez, A. D\'iaz-Guilera,
  J. G\'omez-Garde\~nes, C. J. P\'erez-Vicente, Y. Moreno and
  A. Arenas,  {Phys. Rev. Lett.} \textbf{110}, 028701 (2013).
  
\bibitem{Radicchi}
F. Radicchi and A. Arenas, Nature Physics, {\bf 9},717 (2013).

\bibitem{Perc}
Z. Wang, A. Szolnoki,  M. Perc, EPL,  EPL {\bf 97}, 48001 92012).

\bibitem{Newman1}
J. Park and M. E. J. Newman, Phys. Rev. E {\bf 70}, 066117 (2004).
  
\bibitem{Newman2}
J. Park and M. E. J. Newman, Phys. Rev. E {\bf 70},066146 (2004).
\bibitem{entropyEPL}
G. Bianconi, EPL {\bf 81}, 28005 (2008).
\bibitem{AB2009}
K. Anand and G. Bianconi, Phys. Rev. E {\bf 80}, 045102 (2009).

\bibitem{Garlaschelli}
T. Squartini, G. Fagiolo, D. Garlaschelli, Phys. Rev. E 84, 046118 (2011).
\bibitem{uno}
D. Garlaschelli,New J. Phys. {\bf 11}, 073005 (2009).
\bibitem{due}
T. Squartini, D. GarlaschelliNew J. Phys. {\bf 13}, 083001 (2011).
\bibitem{tre}
D. Garlaschelli and M. Loffredo, Phys. Rev. Lett. {\bf 102}, 038701 (2009).
\bibitem{Multiedge}
O. Sagarra, C. J. P\'erez Vicente, and A. D\'iaz-Guilera,
Phys. Rev. E 88, 062806 (2013).
\bibitem{Multiedge2}
O. Sagarra, F. Font-Clos, C. J P\'erez-Vicente, A. D\'iaz-Guilera, EPL   {\bf 107}, 38002 (2014).
\bibitem{BCoolen}
G. Bianconi, A.C.C. Coolen, C.J.P. Vicente,
Phys. Rev. E 78, 016114 (2008).

\bibitem{Coolen}
A. Annibale, A. C. C. Coolen, L. P. Fernandes, F. Fraternali and J. Kleinjung,
J. Phys. A: Math. Theor. 42 485001 (2009).
\bibitem{delGenio}
C.I. Del Genio, H. Kim, Z. Toroczkai, K.E. Bassler,


\bibitem{Weighted09}
V. Zlatic, G. Bianconi, A. D\'iaz-Guilera, D. Garlaschelli, F. Rao, G. Caldarelli,
Eur. Phys. Jour. B {\bf 67}, 271 (2009).

\bibitem{PNAS}
G. Bianconi, P. Pin and M. Marsili, PNAS {\bf 106}, 11433 (2009).

\bibitem{Barrat_PNAS}
A. Barrat, M. Barth\'elemy, R. Pastor-Satorras and A. Vespignani, PNAS,  {\bf 101} 3747 (2004).

\bibitem{Aalmas}
E Almaas, B Kovacs, T Vicsek, ZN Oltvai, AL Barab\'asi, Nature {\bf 427}, 839 (2004).


\bibitem{ncbi}
www.ncbi.nlm.nih.gov/geo, GSE4183.
\bibitem{kegg}
www.genome.jp/kegg
\bibitem{pathway}
www.pathwaycommons.org
\bibitem{SM}
See Supplementary Material at URL  


\end{thebibliography}
\end{document}